\documentclass[preeprint,nofootinbib,superscriptaddress,floatfix,preprintnumbers]{revtex4-1}
\usepackage{hyperref}
\usepackage{amsmath}
\usepackage{amsfonts}
\usepackage{amssymb}
\usepackage{color}
\usepackage[utf8]{inputenc}
\usepackage{graphicx}
\usepackage{microtype}
\usepackage{siunitx}
\usepackage{subfigure}
\usepackage{soul}
\usepackage[normalem]{ulem}
\usepackage[utf8]{inputenc}
\DeclareUnicodeCharacter{2212}{-}
\usepackage{comment}

\newcommand{\ie}{{\it et al.}}

\begin{document}

\title{Confronting the dark matter capture rate with a continuous gravitational wave probe of local neutron stars}

\author{Pooja Bhattacharjee}\email{pooja.bhattacharjee@ung.si}
\affiliation{Center for Astrophysics and Cosmology, University of Nova Gorica, Vipavska 13, SI-5000 Nova Gorica, Slovenia}
\affiliation{Laboratoire d'Annecy De Physique Des Particules,  CNRS, USMB, F-74940 Annecy, France}
\author{Amit Dutta Banik}\email{amitdbanik@gmail.com}
\affiliation{Physics and Applied Mathematics Unit, Indian Statistical Institute, Kolkata-700108, India}

\begin{abstract}
\noindent Continuous gravitational waves (CGWs) from various astrophysical sources are one of the many future probes of upcoming gravitational wave (GW) search missions. Neutron stars (NSs) with deformity are one of the leading sources of CGW emissions. In this work, for the first time, a novel attempt to estimate the dark matter (DM) capture rate is performed using CGW as the probe to the local NS population. Competitive bounds on DM capture from the local NS population are reported when compared with DM
direct search experiments and other astrophysical observations.
\end{abstract}

\maketitle

\section{Introduction}

\noindent Nearly half a century ago, the existence of dark matter (DM) was confirmed from the observations of spiral galaxies. Later experiments like Planck~\cite{Planck:2018vyg} also verified the presence of DM in the Universe with the observation of cosmic microwave background radiation. However, the nature of DM remains an enigma that needs to be resolved as we are yet to detect a viable DM candidate. Despite its prevalence, the fundamental nature of DM remains unknown, fueling intense research efforts to detect and
understand its properties. Among various DM searches, indirect detection of DM \cite{Slatyer:2017sev} has gained a lot of attention as the terrestrial direct detection experiments failed to detect any promising DM-nucleon scattering event. Indirect detection methods involve searching for the products resulting from annihilation or decay of DM particles ($\chi$), such as gamma rays, neutrinos, or cosmic rays, which can provide indirect evidence of its presence \cite{Gaskins:2016cha}. This detection method has the potential to contribute to the ongoing quest to unravel the enigma of DM and shed light on the fundamental nature of the Universe's underlying structure. \\

\noindent Neutron stars (NSs), the most compact stars in the Universe, emerge from the core-collapse supernova explosions of massive stars \cite{deLavallaz:2010wp}. Their intense gravitational pull and diminutive size render them potent traps for passing DM particles, potentially accumulating them over time \cite{Bramante:2017xlb}. Despite being smaller than brown dwarfs (BDs) \cite{Nakajima:1995, Burrows:2001rb, Leane:2021ihh}, white dwarfs (WDs) \cite{Bell:2021fye, Dasgupta:2019juq}, and stars \cite{Bell:2021pyy, Bell:2021esh}, NSs possess a superior advantage for DM capture owing to their substantial gravitational pull and efficient interaction rate with DM \cite{Busoni:2021zoe, Bell:2021fye, Bell:2020jou}. Their core, primarily composed of tightly packed neutrons, fosters conditions conducive to DM interactions \cite{Press:1985ug, Kouvaris:2007ay}. Moreover, the absence of electromagnetic interference facilitates uninterrupted capture processes, augmenting the efficiency of DM accumulation \cite{deLavallaz:2010wp}. This advantage over BDs and stars positions NSs as promising candidates for indirect DM search \cite{deLavallaz:2010wp}. The accumulation of DM within NSs yields a spectrum of potentially observable outcomes, including detection through the annihilation via long-lived mediator ($\phi$) into neutrinos, gamma rays, or other SM particles that escape the surface of NSs \cite{Okawa:2016wrr, Rothstein:2009pm, Niblaeus:2019gjk, Elor:2015bho}. The DM capture rate in Sun, WDs, BDs, and Galactic NS has been explored in several studies \cite{Leane:2020wob, Leane:2021ihh, Bhattacharjee:2022lts, Acevedo:2023xnu, Maity:2023rez, Bell:2020jou, Bose:2021yhz, Miranda:2022kzs, Bell:2021fye, Chen:2023fgr, Bhattacharjee:2023qfi, Niblaeus:2019gjk, Acevedo:2024ttq, Bose:2022ola}.  \\

\noindent Additionally, DM capture in NSs may induce significant consequences such as kinetic heating \cite{Baryakhtar:2017dbj}, collapsing into black holes \cite{Dasgupta:2020mqg}, alteration of NS merger rates \cite{Busoni:2021zoe}, or modification of gravitational wave signatures from binary NS mergers \cite{Finke:2021eio}. They are also considered as one of the smoking gun signatures of continuous gravitational waves (CGW) \cite{Caride:2019hcv, LIGOScientific:2019yhl}. The rotation of NSs causes them to deform, generating a quadrupole moment that serves as the source of continuous GW emanating from the NS itself \cite{Zimmermann:1979ip, Lasky:2015uia}. Recent studies have pursued the null detection of continuous gravitational waves to impose constraints on their population \cite{Reed:2021scb}, providing an estimation of the local population of NSs near Earth. \\

\noindent In this study, we undertake a comprehensive multiwavelength approach, expanding our inquiry from gamma rays to neutrinos in the pursuit of detecting DM particles confined within NSs. This study aims to scrutinize the DM capture rate, particularly within nearby NSs located approximately 0.6 kpc from Earth, by probing the DM annihilation via long-lived mediator ($\phi$) with space-based or ground-based telescopes. The local NS population inferred from CGW emissions allows us to particularly investigate the cumulative population from around $\sim~10^{5}$ local NSs within 0.6 kpc \cite{Reed:2021scb}. Neutrino and gamma-ray searches offer unique advantages in the quest for DM. Both gamma rays and Neutrinos, being electrically neutral, travel directly from the source location unhindered by the Galactic or extragalactic magnetic field. Neutrino signals, due to their weak interaction, travel unhindered from their sources without deflection or attenuation and can gain insights into dense sources, even those at cosmological distances, which other SM particles cannot reach. Telescopes like Fermi-LAT~\cite{fermi, Ajello_2021}, LHAASO~\cite{LHAASO:2021zta, Conceicao:2023tfb}, CTA~\cite{cta, CTAConsortium:2023ybn}, SWGO~\cite{Conceicao:2023tfb}, etc. search for sources of gamma rays from various celestial objects whereas IceCube~\cite{IceCube:2019cia}, IceCube Gen 2~\cite{IceCube:2021oqo}, TRIDENT~\cite{Ye:2022vbk}, KM3NET~\cite{KM3Net:2016zxf} experiments search for neutrino events from similar astrophysical bodies. \\

\noindent The paper is organized as follows. In Sec.~\ref{source}, we canvass the details of a few local NSs of our interest and nearby NS populations obtained from the CGW searches as primary sources of observation. In Sec.~\ref{sec:formulation}, we brief on the theoretical framework of DM capture and annihilation within NS with long-lived mediators followed by a discussion on present and future experiments of significance to our analysis in Sec.~\ref{detectors}. The bounds on DM scattering cross sections obtained from the analysis with detector sensitivity for local NSs along with the NS population estimate of CGW searches are presented in Sec.~\ref{sec:results}. We summarise the work with concluding remarks in Sec.~\ref{sec:conclusion} and a discussion in Sec.~\ref{sec:Discussion} addressing possible improvements of the results when corrections are implemented on present formalism.

\section{Source Selection}
\label{source}

\noindent In this section, we briefly discuss the local NSs of our interest and introduce the local NS population with continuous gravitational wave searches. 

\subsection{Observed nearby NSs}
\label{ns_obs}

\noindent In this study, we focus on nearby local NSs. We use two criteria to select our candidates: (i) distance from Earth, and (ii) their age. Following Vahdat \ie (2021) \cite{Vahdat:2021dxn}, which provides a list of local NSs, we constrain our selection to those within 0.6 kpc and older than $10^6$ years, resulting in a selection of 8 nearby NSs. For the age criterion, we use the data from the ATNF catalog \cite{atnf}. The positions and characteristics of the selected sources are presented in Table \ref{tab:source_datails}.\\

\noindent Additionally, we adopt an NS mass of 1.4 solar masses (\(M_{\odot}\)) and a radius of 10 kilometers as representative values. These choices are grounded in both observational evidence and theoretical models, providing a robust framework for our analysis. NSs are observed to have masses typically ranging from 1.4-2.9\(M_{\odot}\), but a mass of 1.4\(M_{\odot}\) (or sometimes 1.5\(M_{\odot}\)) is widely used in the literature \cite{Bose:2021yhz, Acevedo:2024ttq} due to its prevalence in nature, particularly in binary NS systems. This is further supported by the gravitational wave event GW170817 \cite{LIGOScientific:2017vwq, Capano:2019eae}, where the NSs involved had masses close to 1.4\(M_{\odot}\). Similarly, while the exact radius of an NS can vary depending on its mass and the equation of state, a radius of 10 kilometers is a well-established approximation \cite{Capano:2019eae, Li:2024ayw}. This value falls within the observed range of 10 to 14 kilometers \cite{Capano:2019eae, Li:2024ayw}, as indicated by x-ray observations and gravitational wave data. Together, these assumptions of a 1.4\(M_{\odot}\) mass and 10-kilometer radius offer a simplified yet realistic model that aligns with current astrophysical understanding.\\

\begin{table*}[htbp]
\centering
\caption{Source Details: Column I: Name of our selected NS;
Column II and III: R.A. and DEC in degree; Column IV: Distance ($d_{\rm NS}$) to the local NSs in kpc; Column V: Mass ($M_{\rm NS}$) in $M_{\odot}$; Column VI: Radius ($R_{\rm NS}$) in km; Column VII: Age in year. We take the parameters of column I-IV from Ref.~\cite{Vahdat:2021dxn}. For mass and radius, we consider the general expected values for NS. For age estimation, we rely on the ATNF catalog~\cite{atnf}.}
\label{tab:source_datails}
\begin{tabular}
{|p{2.2cm}p{1.5cm}p{1.5cm}p{1.5cm}p{1.2cm}p{1.2cm}p{1.7cm}|}
\hline 
\hline
Common Name (ID) & RA (deg) & DEC (deg) & Distance (kpc) & Mass ($M_{\odot}$) & Radius (km)  & Estimated age (yr) \\
\hline
\hline
J0711–6830 & 107.9758 & -68.5132 & 0.11 & 1.4& 10  & $\rm 5.84\times10^{9}$ \\
(Source 1) & & & & &  &  \\
\hline
J0745–5353 & 116.2596 & -53.8561 & 0.57 & 1.4& 10  & $\rm 1.25\times10^{6}$ \\
(Source 2) & & & & &  &  \\
\hline
J0945–4833 & 146.4094 & -48.5540 & 0.35 & 1.4& 10  & $\rm 1.09\times10^{6}$ \\
(Source 3) & & & & &  &  \\
\hline
J0957–5432 & 149.4834 & -54.5344 & 0.45 & 1.4& 10  & $\rm \rm \rm 1.66\times10^{6}$  \\
(Source 4) & & & & &  & \\
\hline
J1000–5149 & 150.1173 & -51.8328 & 0.13 & 1.4& 10  & $\rm \rm 4.22\times10^{6}$ \\
(Source 5) & & & & &  & \\
\hline
J1017–7156 & 154.4639 & -71.9449 & 0.26 & 1.4& 10  & $\rm 1.43\times10^{10}$ \\
(Source 6) & & & & &  & \\
\hline
J1725–0732 & 261.3012 & -7.5498 & 0.20 & 1.4& 10  & $\rm \rm 8.85\times10^{6}$  \\
(Source 7) & & & & &  & \\
\hline
J1755–0903 & 268.7932 & -9.0643 & 0.23 & 1.4& 10  & $\rm 3.9\times10^{6}$ \\
(Source 8) & & & & &  & \\
\hline
\hline
\end{tabular}
\end{table*}

\subsection{Local NS population predicted from CGW}
\label{ns_gw}

\noindent Indirect detection of DM and study of its properties from NSs are based on the gamma-ray and neutrino signal from the sources, focusing mainly on the distribution of the galactic NS population. However, only a very few local NSs are detected making it difficult to constrain the DM-nucleon scattering. As a result, local NSs are generally not considered very significant for the search of DM even though they are expected to be one of the most promising celestial objects to capture DM. This motivates us to look beyond the detection of local NSs with complementary experiments and continuous gravitational wave (CGW) searches provide us with a new window of opportunity. Distinguishing the CGW from local NSs from the background CGW not only leads to the detection of unknown NSs but also helps to predict the local NS distribution in the neighborhood of Earth. Hence, with the help of the local NS population from CGW searches, we expect to obtain a significant improvement in the existing limits on DM capture rate by NSs, especially from the observational perspective. \\

\noindent In this section, we delve into the anticipated count of nearby NSs by exploring CGWs. With an estimated age for the Milky Way of approximately $10^{10}$ years and a Galactic supernovae rate of 1 per century \cite{Diehl:2006cf}, it is plausible that around $N_0 \sim 10^8$ NSs have been generated within our Galaxy to date. However, only a fraction of these, primarily a few thousand, have been identified through electromagnetic searches, largely comprising radio pulsars \cite{Manchester2005aj}. Spinning NSs with deformity in their shapes are promising sources of gravitational waves. The detection of gravitational waves originating from NSs may facilitate the discovery of some of the remaining unidentified NSs and enable the investigation of their morphological distribution \cite{Caride:2019hcv, LIGOScientific:2019yhl}. A recent investigation by Ref. \cite{Reed:2021scb}, conducted an extensive search for CGWs originating from unidentified Galactic NSs to establish constraints on their shapes. This study employed a straightforward model of the spatial and spin distributions of Galactic NSs to estimate the total count of NSs ($N_{\rm NS}$) within proximity to Earth, given a specific upper limit on $\epsilon$. \\

\noindent The emission of gravitational waves from a rotating NS is contingent upon asymmetric deformations within its structure, manifesting as the star's ellipticity ($\epsilon$) \cite{Zimmermann:1979ip, Lasky:2015uia}, defined as
\begin{eqnarray}
\epsilon=\sqrt{\frac{8\pi}{15}}\frac{Q_{22}}{I_{zz}}\, ,
\label{eq:ell}
\end{eqnarray} where quadrupole moment due to deformation is $Q_{22}$, and $I_{zz}$ is the principal moments of inertia along $z$ direction. Considering a rotating NS at a distance $d_{\rm NS}$ and spin frequency $\nu$. The corresponding strain amplitude $h_0$ of the NS is expressed as
\begin{eqnarray}
h_0 = \frac{4\pi^2 G_{\rm N}}{c^4}\frac{I_{zz}f_{\rm GW}^2}{d_{\rm NS}}\epsilon \ ,
\label{epsilon}
\end{eqnarray}
where $f_{\rm GW}=2\nu$, and $G_{\rm N}$ is the gravitational constant. Analysis of CGW search by Abbott \ie \cite{LIGOScientific:2019yhl}, provides a limit of the strain $h_0$ as a function of $f_{\rm GW}$. Using the above information, the expected maximum distance $d_{\rm NS}$ of NS with frequency of rotation $\nu$ and ellipticity $\epsilon$ can be obtained from the nonobservation of any signal from CGW searches. Present GW interferometers can detect signals for emission frequency $f_{\rm GW}>20$ Hz. With the help of spatial distribution of NSs from the detector $N(d_{\rm NS})$, assuming a total of $10^8$ NSs and further considering the spin distribution of observed NSs is the true NS spin distribution in the overall galaxy, an estimate of NS population is obtained in work by Reed et. al. \cite{Reed:2021scb} as a function of ellipticity $\epsilon$ for observable NSs with $f_{\rm GW}>20$ Hz
excluding about 83\% NS population as per observation \footnote{We refrain from the details of the analysis and the complete analysis can be found in the literature \cite{Reed:2021scb}.} The study of NS population density including the unknown population reveals that about $10^5$ local NSs with $\epsilon\simeq 10^{-5}$ within 0.6 kpc distance from Earth. Therefore, the large local population of unknown NSs is expected to improve the bounds on DM properties obtained concerning the observed nearby NSs discussed in Sec. \ref{ns_obs}.
As detailed in Sec. \ref{ns_obs}, our focus is solely on the local NSs within 0.6 kpc from Earth. Following the population model proposed by Ref. \cite{Reed:2021scb} where they showed the possible distribution of NS as a function of $d_{\rm NS}$, the estimated number of NSs within a distance 0.6 kpc from Earth with $\epsilon~\sim~10^{-5}$ is found to be $N_{\rm ns}\simeq 10^{5}$. In subsequent sections, we elaborate on how such a large expected $N_{\rm NS}$ population within 0.6 kpc of Earth can impact our investigation. We consider neutron stars with mass 1.4$M_{\odot}$ with radius 10 km and older than 10$^6$ years satisfying the equilibrium condition for DM capture discussed later in Sec.~\ref{sec:formulation} for the purpose of our analysis\footnote{Similar configuration of neutron stars with 1.5$M_{\odot}$ and radius 10 km was adopted in literature with galactic NS population~\cite{Bose:2021yhz, Acevedo:2024ttq}.}.

\section{Formulation}
\label{sec:formulation}

\noindent This section elucidates the key factors influencing the anticipated gamma-ray and neutrino flux stemming from DM captured within NSs. Our focus is on scenarios where DM interacts with nucleons, resulting in gravitational binding to the NS and subsequent accumulation within it. This occurs when, following single or multiple scatterings, the velocity of the DM particle diminishes below the NS escape velocity, thus preventing evaporation.\\

\noindent The computation can be segmented into three primary parts:\\

\noindent (A) DM capture into NSs, contingent upon the local density and velocity distribution of DM, alongside the scattering cross section of DM particles with neutrons.\\
\noindent (B) The annihilation rate of DM particles ensnared gravitationally within NSs.\\
\noindent (C) The emission of gamma-rays and neutrinos ensuing from the decay of long-lived mediators generated through DM annihilation.

\subsection{DM capture rate inside NSs}
\label{sec:dm_capture}

\noindent The ``maximum capture rate'' ($C_{\rm{max}}$) delineates the scenario where DM particles traversing through NSs become captured. An initial evaluation for $C_{\rm{max}}$ in a DM environment, characterized by parameters such as radius $R_{\rm NS}$, DM number density $n_{\chi}=\rho_\chi/m_\chi$, velocity dispersion ($\bar{v}$), and relative velocity ($v_{0}$), is given by Eq. (\ref{eqn:Cmax}). The NS absorbs the kinetic energy of DM when DM is accelerated to relativistic speeds by its intense gravitational potential. To properly account for the gravitational focusing effect, the escape velocity ($v_{\rm{esc}}$) is modified and expressed as $\sqrt{2 \chi}$, where $ \chi = 1-\sqrt{1-\frac{2G_{\rm N}M_{\rm NS}}{R_{\rm NS}}}$ \cite{Leane:2020wob, Bell:2020jou, Leane:2024bvh}.\\ 

\begin{equation}
C_{\rm{max}}(r) = \pi R_{\rm NS}^{2} n_{\chi}(r) v_{0} \left( 1+ \frac{3}{2} \frac{v^{2}_{\rm{esc}}}{\bar{v}(r)^{2}} \right) \label{eqn:Cmax}
\end{equation}

\noindent Here, $v_{0} = \sqrt{8/(3\pi)}\bar{v}$, and $\bar{v}(r) = 3/2 v_{\rm{c}}(r)$, where $v_{\rm{c}}(r) = \sqrt{G_{\rm N} M(r)/r}$ represents the circular velocity at any galactocentric distance $r$, with $M(r)$ indicating the total Galactic mass encompassed within $r$. Regarding the DM density distribution, we adopt the Navarro-Frenk-White (NFW) density profile \cite{Navarro:1995iw}, as it is widely used in the literature and provides a well-motivated framework consistent with both observational data and the theoretical understanding of DM halos, with $\rho_{\chi(r = R_{\odot})} \equiv \rho_{0} = 0.39$ GeV cm$^{-3}$ \cite{Calore:2022stf}.\\

\noindent In a realistic setting, the perturbative approach must be taken into account, as not all DM particles passing through are captured. As mentioned, DM particles can undergo multiple scatterings, and when their velocity falls below the escape velocity of the NS, they become captured. The probability of experiencing $N$ scatterings before being trapped inside the NS is defined by Eq. (\ref{eqn:p_N}), where $\tau = \frac{3}{2} \frac{\sigma_{\chi n}}{\sigma_{\rm{sat}}}$, with $\sigma_{\chi n}$ denoting the DM-nucleon scattering cross section, $\sigma_{\rm{sat}} = \pi R_{\rm NS}^{2}/N_{n}\simeq 6.58\times 10^{-45}$ cm$^{2}$ \cite{Bramante:2017xlb}, representing the saturation cross section, and $N_{n}=M_{\rm NS}/m_n$ indicating the number of nucleons in the target. As highlighted in recent works \cite{Bell:2020jou, Bell:2020obw, Anzuini:2021lnv}, $N_{n}$ and $v_{\rm{esc}}$ might not be constant and vary across the NS interior due to strong gravitational effects and varying densities. However, in this work, we have adopted a simplified model for direct comparisons and benchmarking of our results within a familiar context \cite{Leane:2024bvh, Acevedo:2024ttq, Leane:2021ihh, Bose:2021yhz}, we assume $N_{n}$ and $v_{\rm{esc}}$ as constants within the NS interior.\\

\noindent The total capture rate ($C_{\rm{tot}}$) encompasses both single and multiple scatterings, as delineated in Eq. (\ref{eqn:c_tot}). \\

\begin{equation}
p_{N}(\tau) = 2,\int_{0}^{1}, dy ,\frac{y,e^{-y\tau},(y\tau)^N}{N!} \label{eqn:p_N}
\end{equation}

\begin{equation}
C_{\rm{tot}}(r) = \sum_{N=1}^{\infty} C_{N}(r) \label{eqn:c_tot}
\end{equation}

\noindent The capture rate corresponding to $N$ scatterings, $C_{N}$, is delineated by Eq. (\ref{eqn:c_N}) \cite{Bramante:2017xlb,Ilie:2020}. Here, $v_{N}$ is defined as $v_{N} = v_{\rm{esc}}(1-\beta_{+}/2)^{-N/2}$, where $\beta_{+} = 4m_\chi m_{n}/(m_\chi+m_{n})^{2}$. For a large number of scatterings ($N \gg 1$), ($v_{N}^{2} - v_{\rm{esc}}^{2}$) predominates over $\bar{v}^{2}$, and $C_{N}$ approximates $\approx p_{N} C_{\rm{max}}$.

\begin{equation}
C_{N}(r) = \frac{\pi R_{\rm NS}^{2} p_{N}(\tau)}{1-2GM_{\rm NS}/R_{\rm NS}} \frac{\sqrt{6}n_{\chi}(r)}{3\sqrt{\pi}\bar{v}(r)} \left[(2\bar{v}(r)^{2} + 3v_{\rm{esc}}^{2}) - (2\bar{v}(r)^{2} + 3v_{N}^{2})\exp\left(-\frac{3(v_{N}^{2} - v_{\rm{esc}}^{2})}{2\bar{v}(r)^{2}}\right)\right] 
\label{eqn:c_N}
\end{equation}

\noindent We emphasize that for $m_{\chi} < 1$ GeV, the accretion rate no longer scales as $1/m_{\chi}$ but is instead suppressed due to Pauli blocking (PB). We scale the $\sigma_{\rm{sat}}$ value for $m_{\chi} < 1$ GeV by following the Ref.~\cite{Baryakhtar:2017dbj}. As reported in some recent studies \cite{Bell:2020jou, Garani:2018kkd}, we analyze the changes in $C_{\rm{tot}}$ for $m_{\chi} < 1$ GeV in comparison to $C_{\rm{tot}}$ at $m_{\chi} = 1$ GeV. Following the Pauli blocking suppression reported in Ref.~\cite{Bell:2020jou} with respect to $C_{\rm{tot}}$ at $m_{\chi} = 1$ GeV, we define a scaling factor $f_{\rm PB}$ which is $\sim 1.6 \times m_{\chi}^{1.2}$ valid up to $ m_{\chi} < 1$ GeV \footnote{It is important to note that the $\sigma_{\rm{sat}}$ values for the selected sources are comparable to the geometric limits used in Ref.~\cite{Bell:2020jou}. As a result, our scaling formula provides a reasonably accurate estimate of the expected Pauli blocking effect, avoiding significant overestimation or underestimation.} By applying this scaling factor to Eq.~(\ref{eqn:c_tot}), we determine the capture rate under Pauli blocking suppression, $C_{\rm{PB}}(r)$, for $m_{\chi} < 1$ GeV, given by,

\begin{equation}
C_{\rm{PB}}(r) = C_{\rm{tot}}(r) \times f_{\rm PB} 
\label{PB_capture}
\end{equation}

\subsection{DM annihilation rate}
\label{sec:dm_annihilation}

\noindent In this section, we discuss the DM annihilation rate resulting from accumulated DM inside NS. We want to point out that for this study we ignore the impact of DM evaporation, effective for $m_{DM}\leq$ 10 eV for a Gyr old NS\cite{Bell:2020lmm, Bell:2023ysh}. Therefore, in the absence of evaporation, the total number of DM particles, represented as \(N(t)\), accumulated within the core of the NS at time \(t\), is governed by the evolution of DM number abundance in Eq.~(\ref{eqn:N_variation}):

\begin{equation}
\frac{{\rm d}N(t)}{{\rm d}t} = C - C_{\rm ann}N^2(t)\,, \\
\label{eqn:N_variation}
\end{equation}

\noindent In this context, $C = \min[C_{\rm{tot}}, C_{\rm{max}}]$ denotes the total capture rate under the condition that the perturbative estimation is deemed valid, while $C_{\rm ann}$ signifies the annihilation rate defined as $C_{\rm ann}=\frac{\langle \sigma v \rangle}{V_{\rm NS}}$; where $\langle \sigma v \rangle$ represents the annihilation cross section for thermal DM and $V_{\rm NS}$ denotes the volume within the NS where annihilation occurs. As we disregard the influence of evaporation, the solution to Eq. (\ref{eqn:N_variation}) can be expressed as:

\begin{equation}
N(t) = C\,t_{\rm eq} \tanh\frac{t}{t_{\rm eq}} \label{eqn:Neq_time}
\end{equation}

\noindent The equilibrium time, $t_{\rm eq}~\equiv \frac{1}{\sqrt{C_{\rm ann}C}}$, represents the time required to reach equilibrium between the DM capture rate and the annihilation rate in the absence of evaporation. If equilibrium is achieved today (i.e., $t_{\rm NS} \ge t_{\rm eq}$), the total annihilation rate ($\Gamma_{\rm ann}$) is solely dependent on the DM capture rate, i.e., $\Gamma_{\rm ann} \rightarrow \frac{C}{2}$. As mentioned in Ref.~\cite{Bell:2023ysh}, the equilibrium age of an NS is under one year for vector interactions and approximately $10^{4}$ years for scalar interactions. Given that all our sources are older than $10^{6}$ years, we can conclude that the equilibrium between capture and annihilation rates has been established.

\subsection{Gamma-ray and neutrino spectrum: For DM annihilation via long-lived mediators}
\label{sec:dm_spectrum}

\noindent In this investigation, we aim to probe the gamma-ray and the neutrino emission resulting from the DM annihilation inside NS. Typically, when DM particles undergo direct annihilation into quark-antiquark pairs, they expect to generate detectable signals in the form of gamma rays, neutrinos, or any other form of SM products as final states. However, when examining DM annihilation within dense celestial bodies such as NSs, there exists a possibility that if DM particles directly annihilate into SM final states, they might become confined within the NS, thereby inducing its heating. This heating phenomenon could potentially be observed using the James Webb Space Telescope (JWST), as recently discussed in reference~\cite{Leane:2021ihh}, although such observations pose significant challenges due to current instrumental limitations. Nevertheless, the chances of detecting a signal are enhanced when DM annihilation to SM states occurs via long-lived mediators ($\phi$) that can escape the NS's surface, as proposed in various references \cite{Pospelov:2007mp, Pospelov:2008jd, Batell:2009zp, Dedes:2009bk, Fortes:2015qka, Okawa:2016wrr, Yamamoto:2017ypv, Holdom:1985ag, Holdom:1986eq, Chen:2009ab, Rothstein:2009pm, Berlin:2016gtr, Cirelli:2016rnw, Cirelli:2018iax}. The decay products of these mediators may potentially yield observable signals in gamma-ray or neutrino telescopes. \\

\noindent Nonetheless, there remains a possibility that these mediators might interact with SM constituents within the NS before escaping, potentially resulting in a significant decrease in the observed flux of gamma rays and neutrinos. As illustrated in \cite{Leane:2021ihh, Bhattacharjee:2022lts}, it is possible to substantially mitigate this attenuation by carefully selecting appropriate model parameters. Hence, for the sake of simplicity and ease of comparison with existing literature, we focus on the scenario where the long-lived mediator facilitates DM annihilation outside the NSs, such as $\chi \chi \rightarrow \phi \phi \rightarrow 4 \gamma$ and $\chi \chi \rightarrow \phi \phi \rightarrow \nu\bar{\nu}\nu\bar{\nu}$. Additionally, we make the assumption that \(m_\phi\ll m_\chi\), under which the provided formulas remain applicable even though this approximation removes the dependence on the additional parameter $m_\phi$. It is to be noted that, we consider DM annihilates into photon or neutrino only with branching ratio unity, and has no other annihilation channel. This corresponds to the maximum flux obtained from DM into the specific annihilation channel. In general, DM can have many annihilation channels depending on the mass. With the specific choice of models where DM is neutrinophilic, and the mediator $\phi$ is coupled to neutrino only, this kind of DM annihilation scenario can be achieved. However, constructing a complete model for DM is beyond the scope of this work.\\

\noindent The Eqs. (\ref{eqn:dm_flux_gamma}) and (\ref{eqn:dm_flux_nu}) describe the differential flux of gamma rays and neutrinos reaching Earth as a result of DM annihilation via a long-lived mediator.
\begin{equation}
\frac{d \phi_{\gamma}}{d E_{\gamma}} = \frac{\Gamma_{\rm ann}}{4 \, \pi \, d_{\rm NS}^{2}} \times \left(\frac{d N_{\gamma}}{d E_{\gamma}} \right)  \times  P_{\rm surv},
\label{eqn:dm_flux_gamma}
\end{equation}

\begin{equation}
\frac{d \phi_{\nu}}{d E_{\nu}} = \frac{\Gamma_{\rm ann}}{4 \, \pi \, d_{\rm NS}^{2}} \times \left(\frac{d N_{\nu}}{d E_{\nu}} \right)  \times  P_{\rm surv},
\label{eqn:dm_flux_nu}
\end{equation}

\noindent In this context, we take into account the contribution of the ``survival probability'' ($P_{\rm surv}$) for gamma rays and neutrinos reaching the detectors, which is nearly unity within the range of mediator decay lengths. In our scenarios, we are considering neutrinos arriving from distances of a few kpc, allowing us to confidently neglect the neutrino oscillation effects and proceed with the assumption that the proportion of neutrino flavors reaching Earth follows a ratio of $\nu_e:\nu_{\mu}:\nu_{\tau}=1:1:1$.\\

\noindent Under our stated hypothesis, where $m_{\phi}\ll m_\chi$, the reliance on $m_{\phi}$ is removed, leading to the conceptualization of the neutrino and gamma-ray spectrum as a box-shaped distribution, as described by Ref. \cite{Ibarra:2012dw} in Eq. \ref{eq:box_function}. 

\begin{equation}
\frac{{\rm d}N_{\gamma,\nu}}{{\rm d}E_{\gamma,\nu}} = \frac{4\Theta(E-E_{-}) \Theta(E_{+}-E)}{\Delta E}\,,
\label{eq:box_function}
\end{equation}

\noindent Here, $\Theta$ signifies the Heaviside-theta function. The energy bounds are denoted by $E_{\pm}=(m_{\chi}\pm \sqrt{m_{\chi}^2-m_{\phi}^2})/2$, with the width of the box function defined as $\Delta E=\sqrt{m_{\chi}^2-m_{\phi}^2}$.

\section{Current and Future generation telescopes}
\label{detectors}

\noindent Studying the indirect detection of DM signals through multiwavelength observations, particularly in gamma-ray and neutrino wavelengths, offers a promising avenue to unravel the mysteries of DM. Gamma rays and neutrinos are two fundamental messengers that can provide complementary insights into the elusive nature of DM particles. Gamma rays, emitted from astrophysical sources or produced by DM annihilation or decay, can unveil signatures indicative of DM interactions. On the other hand, neutrinos, being weakly interacting particles, can traverse vast cosmic distances without being absorbed or deflected, offering a pristine view of the Universe's most distant and obscured regions.\\

\noindent Combining data from gamma-ray and neutrino observations enables us to construct a comprehensive picture of DM distribution and behavior across different cosmic scales. Moreover, cross correlating these observations with other astronomical datasets, such as those from gravitational wave detectors and galaxy surveys, enriches our understanding of DM's role in shaping the cosmos. Multiwavelength studies thus stand as a powerful strategy to decipher the enigmatic properties of DM and elucidate its profound impact on the Universe.\\

\noindent In Sec.~\ref{sec:dm_spectrum}, we discuss the shape of gamma-ray and neutrino spectra resulting from DM annihilation through long-lived mediators. In this section, we probe the expected gamma-ray and neutrino flux with the current and future generations of neutrino and gamma-ray telescopes.

\subsection{With current and future neutrino telescopes}
\label{neutrino_detector}
\noindent Neutrino telescopes offer a novel avenue for probing DM \cite{IceCube:2021xzo, Miranda:2022kzs} by detecting neutrinos. These telescopes complement traditional detection methods, providing insight into the distribution and properties of DM through neutrino emissions from potential DM sources throughout the Universe. \\

\noindent IceCube, a massive cubic-kilometer neutrino detector situated at the South Pole, was completed in 2010. Originally proposed based on theoretical calculations in 1998, it was envisioned to be sensitive enough to detect high-energy neutrinos from phenomena such as AGN jets or GRBs \cite{Waxman:1998yy}. As the first experiment of its kind, IceCube has achieved significant milestones, including the discovery of a diffuse extraterrestrial neutrino flux in 2013 \cite{IceCube:2013low} and the detection of neutrino emission from a flaring blazar in 2017 \cite{IceCube:2018cha, IceCube:2018dnn}. Despite dedicated efforts to identify the sources of cosmic neutrinos through various analyses, such as all-sky survey \cite{IceCube:2019cia}, transient \cite{IceCube:2022rlk, IceCube:2017fpg, IceCube:2019acm}, and AGN catalog stacking searches \cite{IceCube:2021pgw, IceCube:2016qvd}, conclusive results remain elusive. This suggests the possibility of multiple weaker sources contributing to the diffuse flux \cite{Murase:2015xka}, such as starburst galaxies \cite{Ambrosone:2021brr}, which would necessitate improved pointing resolution \cite{Fang:2016hop}.\\

\noindent IceCube Gen 2, an extension of the IceCube Neutrino Observatory in Antarctica, aims to explore the high-energy neutrino sky from TeV to PeV energies \cite{IceCube:2019pna, IceCube-Gen2:2023vtj}. It promises five times better sensitivity for neutrino sources compared to the current IceCube detector. With 120 new strings and larger spacing between them, IceCube Gen 2 will enhance sensitivity for neutrinos above 10 TeV and target detection of neutrinos exceeding 100 PeV \cite{IceCube-Gen2:2023vtj}. The ongoing IceCube upgrade will improve detection thresholds down to 1 GeV, enhancing capabilities for various physics studies \cite{IceCube:2021oqo}. IceCube Gen 2's larger instrumented area will enable the detection of fainter neutrino sources, positioning it as the world's leading neutrino observatory. It holds promise for diverse scientific inquiries, including DM searches, particle physics studies, and the detection of supernova neutrinos \cite{Ishihara:2019aao}.\\

\noindent TRIDENT, a next-generation neutrino telescope slated for construction in the South China Sea \cite{Ye:2022vbk}, is poised to transform our understanding of high-energy astrophysical neutrinos. Its primary goal is to detect multiple sources and enhance measurements of cosmic neutrino events across all flavors. With improved angular resolution and sensitivity \cite{Hu:2021jjt}, TRIDENT aims to pinpoint sources within the diffuse flux identified by IceCube \cite{Ye:2022vbk}. Positioned near the equator, TRIDENT will provide extensive visibility of the neutrino sky due to Earth's rotation, bolstering the global network of neutrino observatories \cite{Ye:2022vbk}. Furthermore, the reduced light scattering in water compared to IceCube's glacial ice promises enhanced neutrino-pointing accuracy. Together, these advancements are expected to significantly increase the detection of astrophysical neutrinos, paving the way for source identification and advancing the field of neutrino astronomy \cite{Ye:2022vbk, IceCube:2017roe}.\\

\noindent Several telescopes currently in development, including KM3NeT in the Mediterranean Sea \cite{KM3Net:2016zxf}, Baikal-GVD in Lake Baikal \cite{Avrorin:2011zzc}, and the proposed P-ONE in the East Pacific \cite{P-ONE:2020ljt}, aim to complement IceCube's coverage of the TeV-PeV neutrino sky from the northern hemisphere. KM3NeT is a network of deep-sea neutrino telescopes planned for deployment in the Mediterranean Sea \cite{KM3Net:2016zxf}. The KM3NeT/ARCA detector, located at the Capo Passero site in Italy, is specifically designed to detect high-energy cosmic neutrinos. KM3NeT/ARCA, with a broader field of view than IceCube, focuses on detecting Galactic sources, especially those observable at lower energies around tens of TeV. However, the observatory's sensitivity to muon neutrinos is limited \cite{KM3NeT:2018wnd, KM3NeT:2023kqy}.

\subsection{With current and future gamma-ray telescopes}
\label{gamma_detector}

\noindent Multimessenger astrophysics has become a groundbreaking reality, propelled by recent advancements in gravitational wave, gamma-ray astronomy, and high-energy neutrino detection. These breakthroughs have ushered in a new era of understanding the high-energy astrophysical universe and the mechanisms underlying its most energetic phenomena. \\

\noindent Over the past two decades, advancements in both space-based and ground-based gamma-ray detectors have greatly enhanced our understanding of the high-energy gamma-ray universe. At high energies (HE, E $>$ 0.1 GeV), the Fermi Large Area Telescope (Fermi-LAT) continues to be a linchpin in multiwavelength and multimessenger studies \cite{Ajello_2021}. Surveying the gamma-ray sky since 2008, Fermi-LAT's comprehensive energy coverage and expansive field of view have facilitated groundbreaking discoveries across eight orders of magnitude in photon energy \cite{Atwood_2009}. All $\gamma$-ray data from Fermi are made publicly available in real-time, fostering collaborative research across diverse scientific communities. \\

\noindent Looking ahead, for Very High Energy (VHE, E $>$ 0.1 TeV) observations, the next-generation ground-based gamma-ray detector, Cherenkov Telescope Array (CTA), represents a monumental leap in ground-based gamma-ray astronomy \cite{CTAObservatory:2022mvt}. Encompassing an energy range from some tens of GeV to about 300 TeV, CTA's three arrays of Imaging Atmospheric Cherenkov telescopes promise unprecedented sensitivity and precision \cite{Gueta:2021vrf}. With a focus on extragalactic objects in the Northern Hemisphere and galactic sources in the Southern Hemisphere, CTA's vast effective area and field-of-view position it as a cornerstone instrument for future gamma-ray astronomy endeavors \cite{Gueta:2021vrf}. \\

\noindent As the next-generation gamma-ray telescope to observe very- to ultrahigh-energy gamma rays, the Southern Wide-field Gamma-ray Observatory (SWGO), a water Cherenkov telescope array, stands at the forefront of this revolution\cite{Conceicao:2023tfb}. Spanning energies from about 30 GeV to a few PeV, SWGO provides a wide field and high-duty cycle view of the southern sky \cite{Conceicao:2023tfb}, complementing existing particle arrays in the Northern Hemisphere like Large High Altitude Air Shower Observatory (LHAASO) \cite{LHAASO:2021zta}. Speaking of which, LHAASO, is a multipurpose facility designed to study cosmic rays and gamma rays across a broad energy spectrum, ranging from sub-TeV to beyond 1 PeV \cite{Cao:2010zz}.

\section{Results}
\label{sec:results}
\subsection{Differential flux sensitivity}
\label{sec:diff_flux}

\begin{figure}[htbp]
\centering
\includegraphics[width=0.485\linewidth]{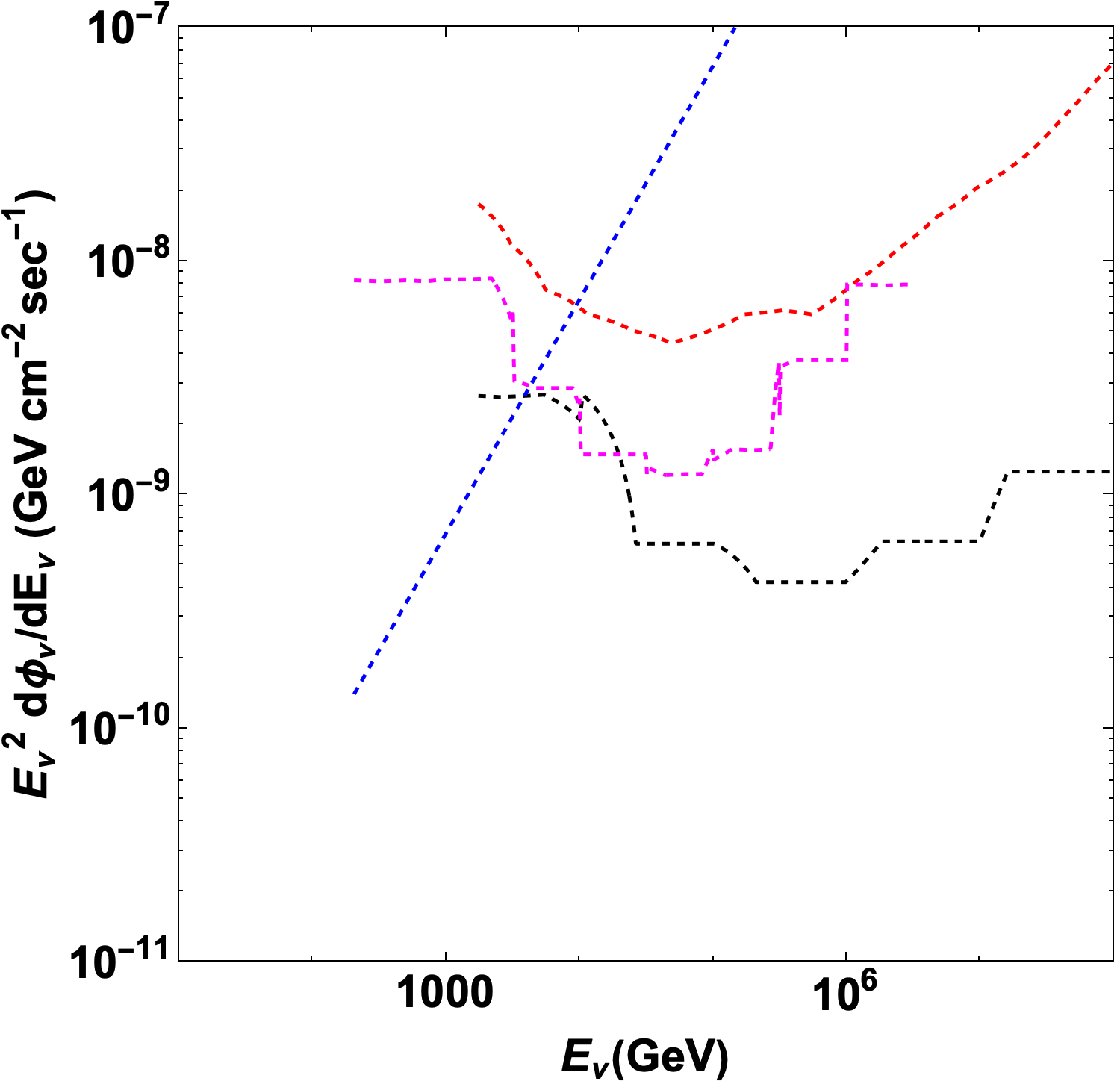}
\includegraphics[width=0.485\linewidth]{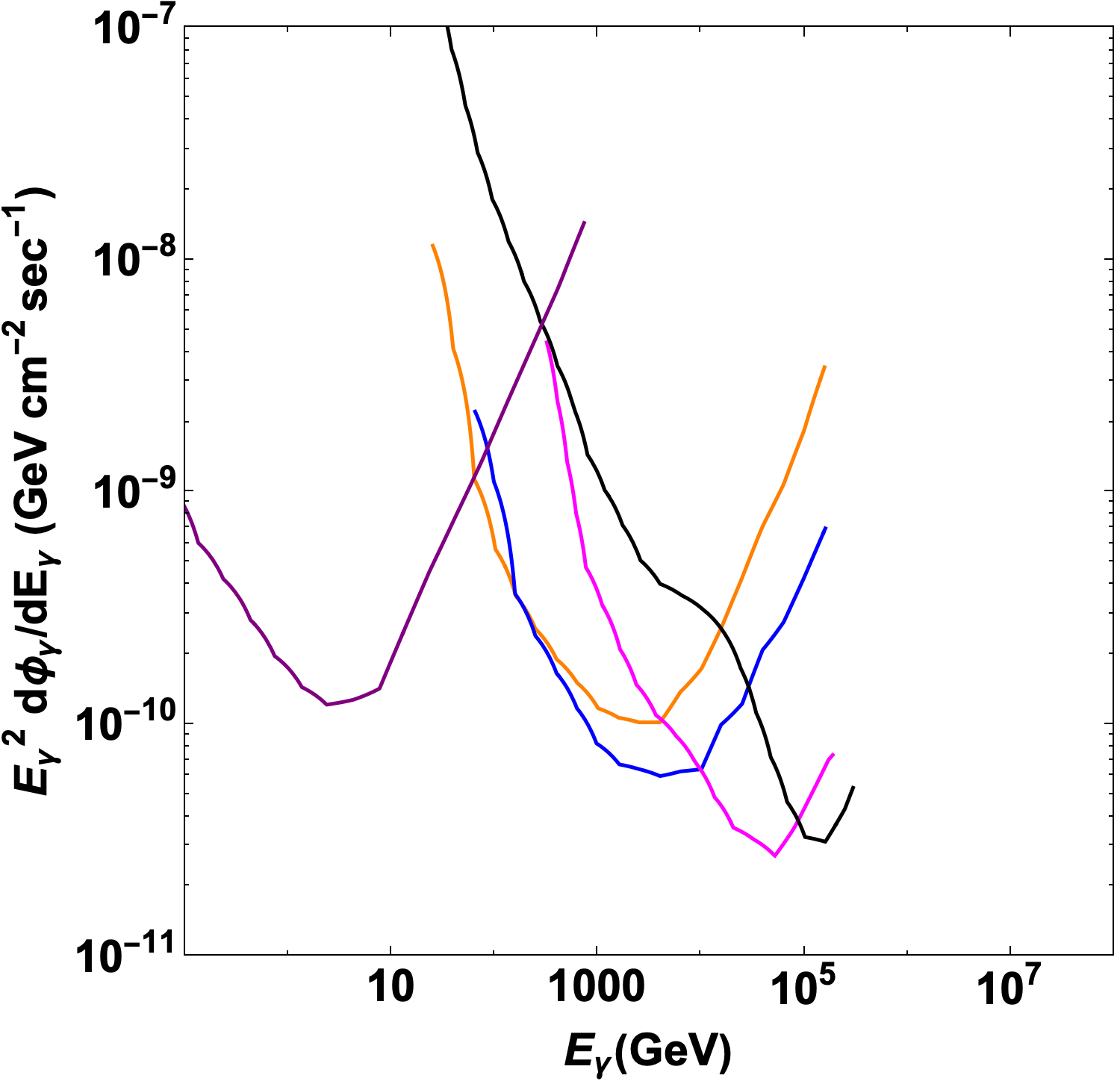}
\includegraphics[width=0.18\linewidth]{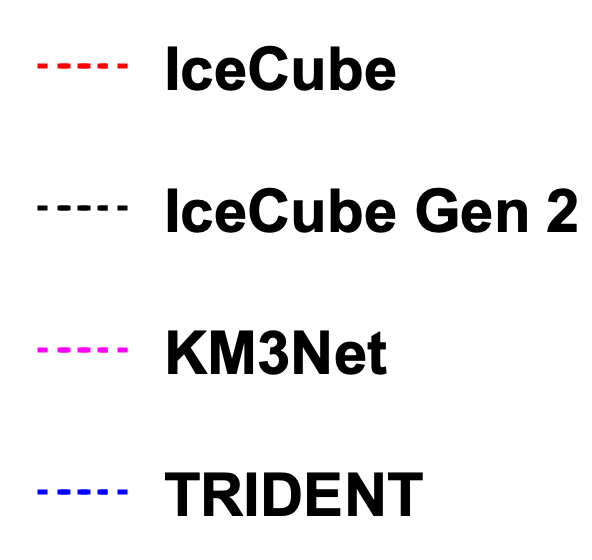}
\includegraphics[width=0.14\linewidth]{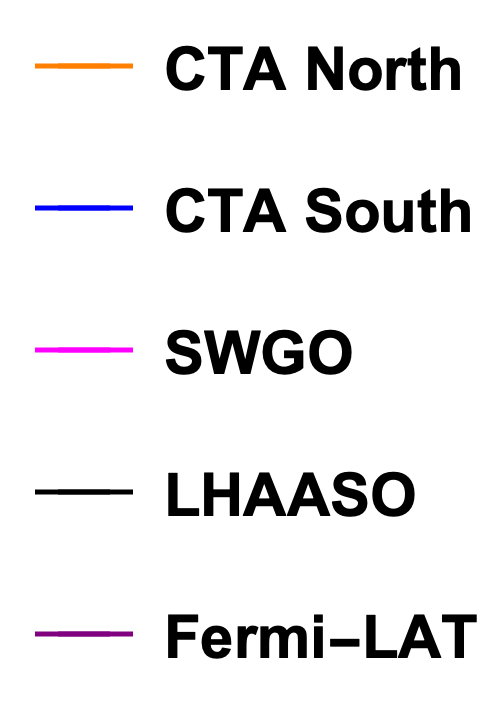}
\caption{Differential flux ($E_{\nu, \gamma}^{2}~d\phi_{\nu, \gamma}/dE_{\nu, \gamma}$) sensitivity of all our selected neutrino (left) and gamma-ray (right) detectors.} 
\label{fig:diff_flux_ref}
\end{figure}
\noindent Following the discussions of neutrino and gamma-ray detectors in Sec.~\ref{detectors}, in this section, we highlight the differential flux sensitivity of these experiments toward the detection of point sources.  
In Fig.~\ref{fig:diff_flux_ref} (Left: neutrino and Right: gamma-ray experiments), we show the differential flux sensitivity of all our selected telescopes mentioned in Sec.~\ref{detectors}. For IceCube and IceCube Gen 2, we take into account the detection sensitivity (as detailed in Ref. \cite{IceCube:2019pna}) to observe the neutrino flux originating from a point source positioned at the celestial equator ($\delta = 0^0$), achieving an average significance of 5$\sigma$ over a decade of observations. As par the planned upgradation, IceCube Gen 2 aims to be sensitive for a very wide energy range starting from 1 GeV to 100 PeV \cite{IceCube:2021oqo} and would become one of the world’s leading atmospheric and astrophysical neutrino observatories. For KM3Net, we consider the 90\% confidence level (C.L.) quasidifferential sensitivity for pointlike emission predicted for the full detector array \cite{KM3NeT:2023kqy}. The potential of the full KM3NeT is expected to discover diffuse and pointlike spectra that have not been detected by IceCube yet. \\

\noindent Finally, for TRIDENT which is expected to be the most sensitive neutrino telescope in the upcoming days, we use the 90\% C.L. median sensitivity for the pointlike sources from the expected all-sky survey \cite{Ye:2022vbk}. In this context, we want to mention that for all neutrino telescopes, we only consider their sensitivity from tracklike events for the sources near declination, $\delta = 0^0$.\\

\noindent For the gamma-ray telescopes, such as CTA (\cite{cta, CTAConsortium:2023ybn}), SWGO (\cite{Conceicao:2023tfb}), LHAASO (\cite{LHAASO:2021zta, Conceicao:2023tfb}) and Fermi-LAT (\cite{fermi, Ajello_2021}), we use the differential flux limits for pointlike sources from their respective performance guidelines.

\subsection{Variation of differential flux sensitivity of neutrino telescopes with declination}
\label{sec:flux_dec}

\begin{figure}[htbp]
\centering
\includegraphics[width=0.48\textwidth]{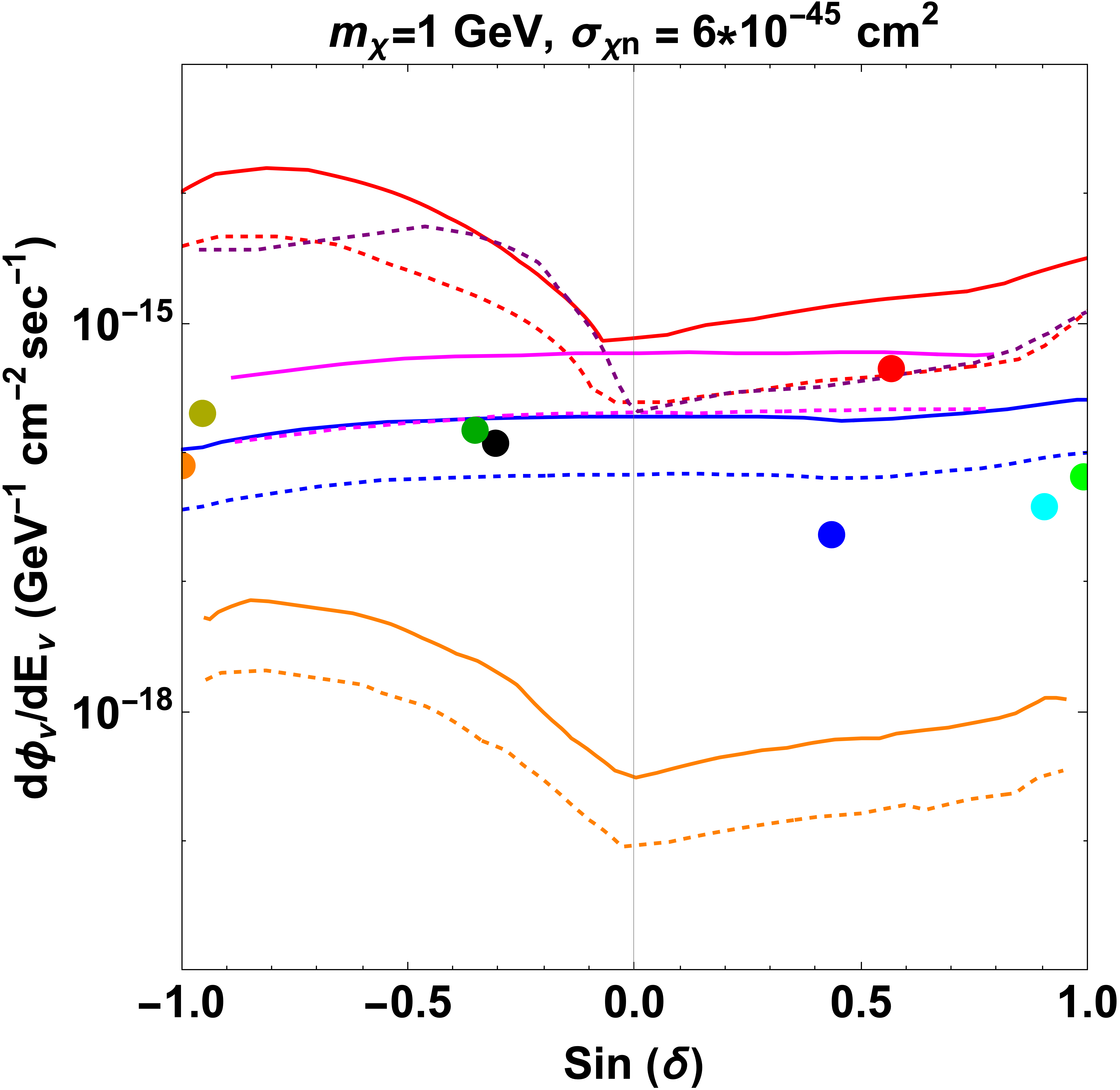}
\includegraphics[width=0.48\textwidth]{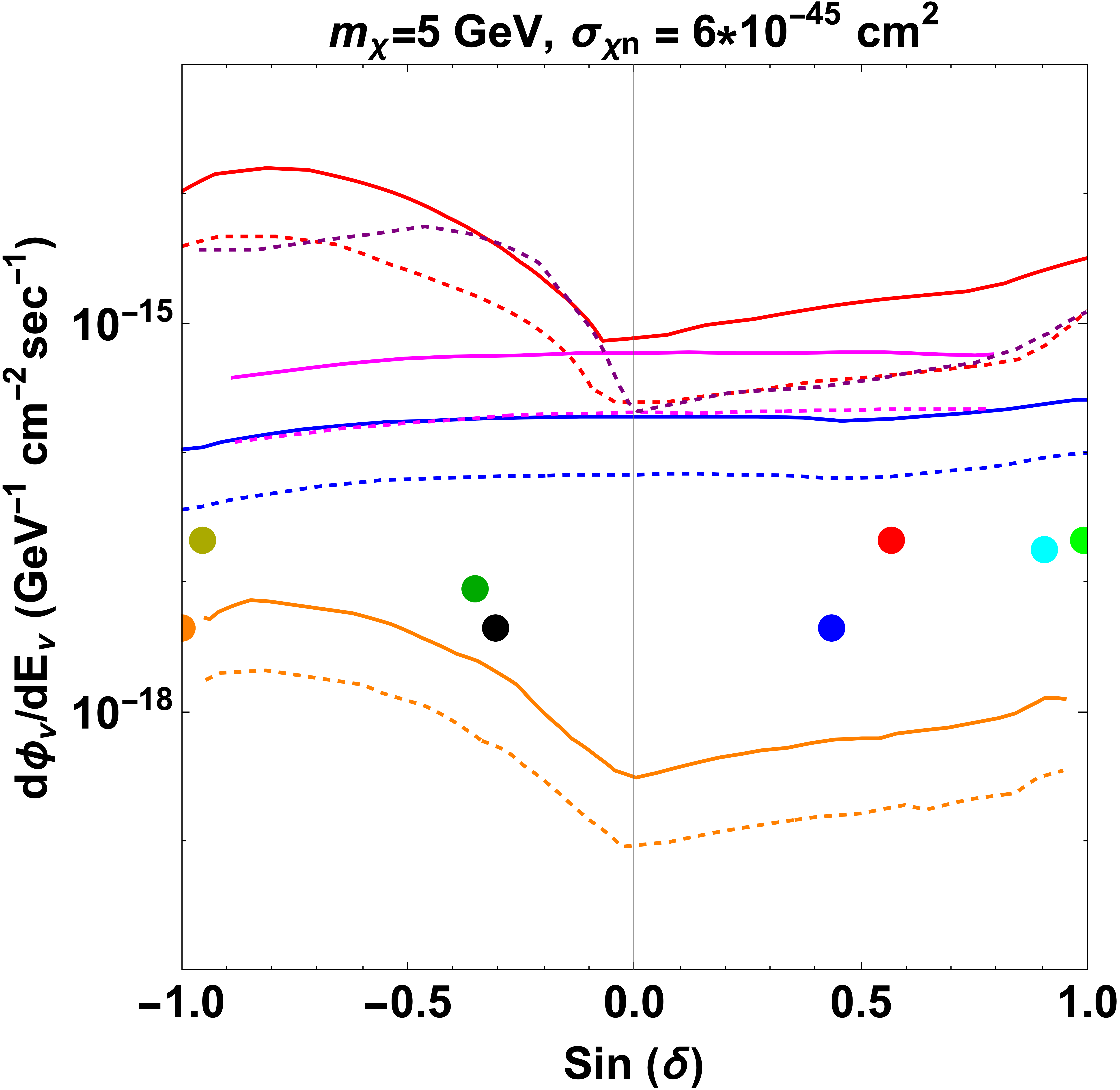}
\includegraphics[width=0.3\textwidth]{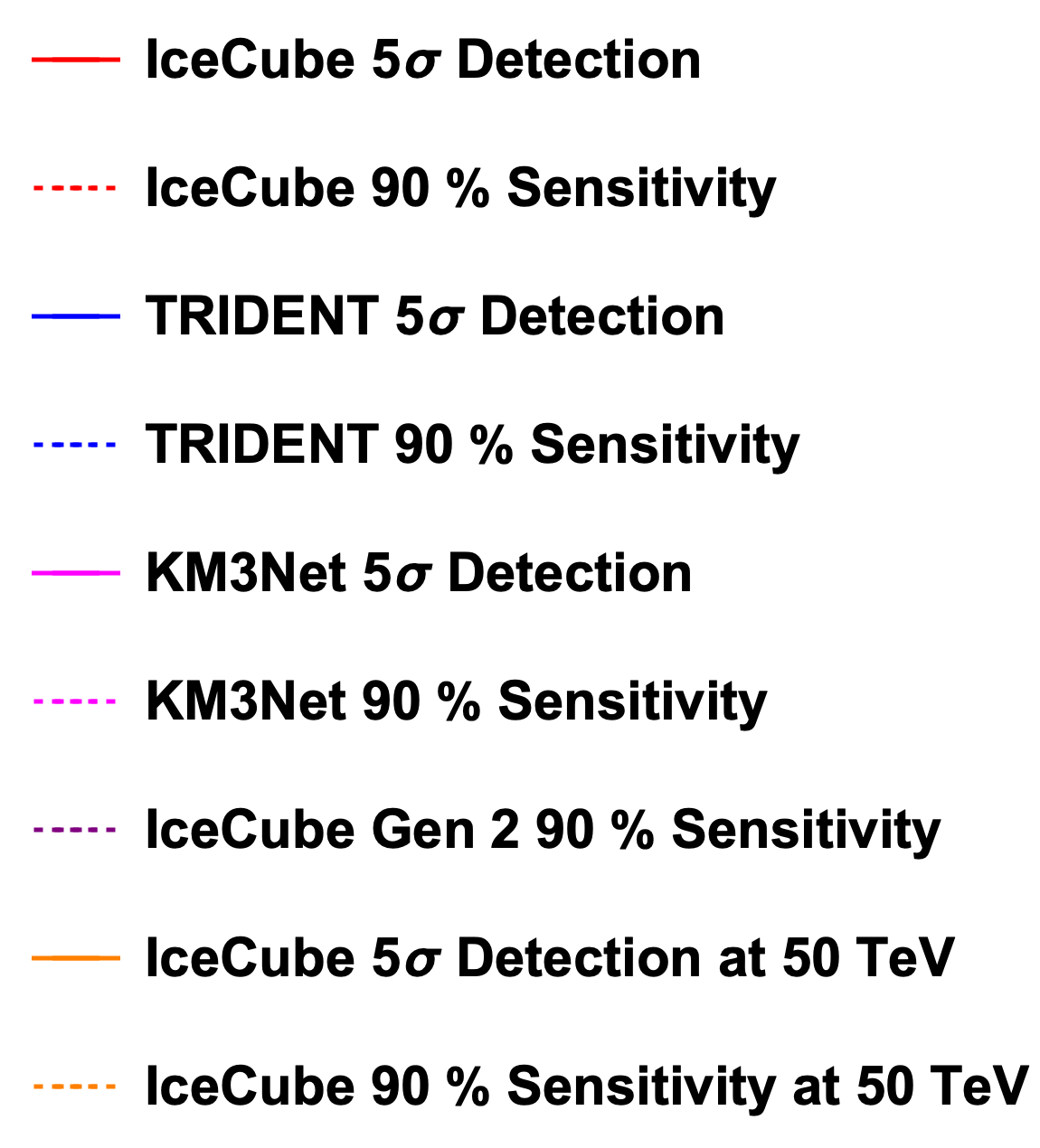}
\includegraphics[width=0.1\textwidth]{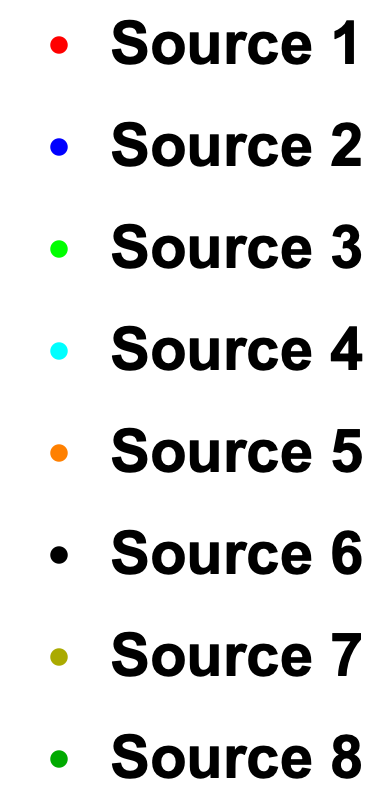}
\caption{Variation of the differential flux of current and future neutrino telescopes with declination for two DM mass values.} 
\label{fig:diff_flux_dec}
\end{figure}

\noindent In this section, we study the neutrino signals from NSs resulting from the decay of long-lived mediators produced from DM annihilation in the light of current and future-generation neutrino telescopes. This study specifically focuses on the DM-nucleon scattering cross section in equilibrium, aiming to explore whether NSs offer a promising avenue for probing small DM-nucleon cross sections. Throughout our investigation, we assume that DM number abundance has reached equilibrium, given the age of NSs considered to be higher ($\ge 1$ Myr) than $t_{\rm eq}$. All of our selected neutrino telescopes are equally sensitive for both spin-independent (SI) and spin-dependent (SD) DM-nucleon scattering cross sections as NSs primarily consist of neutrons. \\

\noindent Fig.~\ref{fig:diff_flux_dec} offers a comprehensive view of the 90\% C.L. sensitivity (depicted by the dashed line) and the 5$\sigma$ discovery potential (shown by the solid line), considering a source spectrum with a differential flux $\frac{dN}{dE}\propto\rm E^{−2}$ as a function of source declination (sin $\delta$)\footnote{When normalized over a comparable energy range, the total flux from a power-law spectrum can be similar to that of a boxlike spectrum, especially over broad energy ranges. This makes comparisons relevant for the sensitivity derived from neutrino telescopes.}. The `points' on the graph represent individual differential neutrino flux values for neutrino flux from each NS for our chosen DM mass ($m_{\chi}$) and scattering cross section ($\sigma_{\chi n}$). \\

\noindent Fig.~\ref{fig:diff_flux_dec} indicates that for IceCube \cite{IceCube:2019cia} and IceCube Gen 2\cite{IceCube-Gen2:2021tmd}, the highest sensitivity is concentrated around the equator, corresponding to optimal discovery potential, while it is weakest for declination values near sin($\delta$) = $\pm 1$ \cite{IceCube:2019cia}. KM3NeT (with 6 years of projected data) \cite{KM3NeT:2018wnd} and TRIDENT \cite{Ye:2022vbk}, provide equal sensitivity for all the regions of the sky. This underscores the challenge of IceCube and IceCube Gen 2 of detecting events of interest amidst the background, particularly in regions where a stronger signal is required. For a source located in the southern sky, TRIDENT will have nearly 4 orders of magnitude improvement in sensitivity \cite{Ye:2022vbk} compared to IceCube. \\

\noindent Consequently, given the current sensitivity of all these neutrino detectors, TRIDENT has the potential to detect local NSs for DM mass around 1 GeV with $\sigma_{\chi n}$ = $6~\times~10^{-45}$ cm$^{2}$.  However, for DM mass for $m_{\chi}~\sim$ 5 GeV, as shown in Fig.~\ref{fig:diff_flux_dec}, the current and future telescopes are not sufficient unless brighter sources are present or DM-nucleon scattering is enhanced further. The choice of $\sigma_{\chi n}$ = $6~\times~10^{-45}$ cm$^{2}$ is motivated by its proximity to the threshold cross section i.e., $\sigma_{\rm{sat}} \simeq 6.58\times 10^{-45}$ cm$^{2}$ \cite{Bramante:2017xlb} of our designated local NSs \footnote{As we use the same mass and radius for each NS, we obtain the same $\sigma_{\rm{sat}}$ value. If $\sigma_{\chi n}$ is less than or equal to $\sigma_{\rm{sat}}$, the DM is expected to lose its kinetic energy and be captured after a single collision. However, once this threshold is exceeded, the effect of multiple scatterings must be considered. To simplify the discussion, in Fig.~\ref{fig:diff_flux_dec}, we focus on the single-scattering scenario.}.

\subsection{Expected bounds on $\sigma_{\chi n}$ from local NSs}
\label{sec:bounds}

\noindent In this section, we aim to estimate the bounds on $\sigma_{\chi n}$ from the gamma-ray and neutrino emission resulting from DM annihilation from the local NSs within 0.6 kpc from Earth. First, we look for the emission from our selected 8 local NSs, and next, we study how the expected number of local NSs predicted from CGW boosts the current bounds. 

\subsubsection{Limits from eight local NSs with neutrino telescopes}
\label{sec:limits_nu_only}

\begin{figure}[htbp]
\centering
\includegraphics[width=0.65\textwidth]{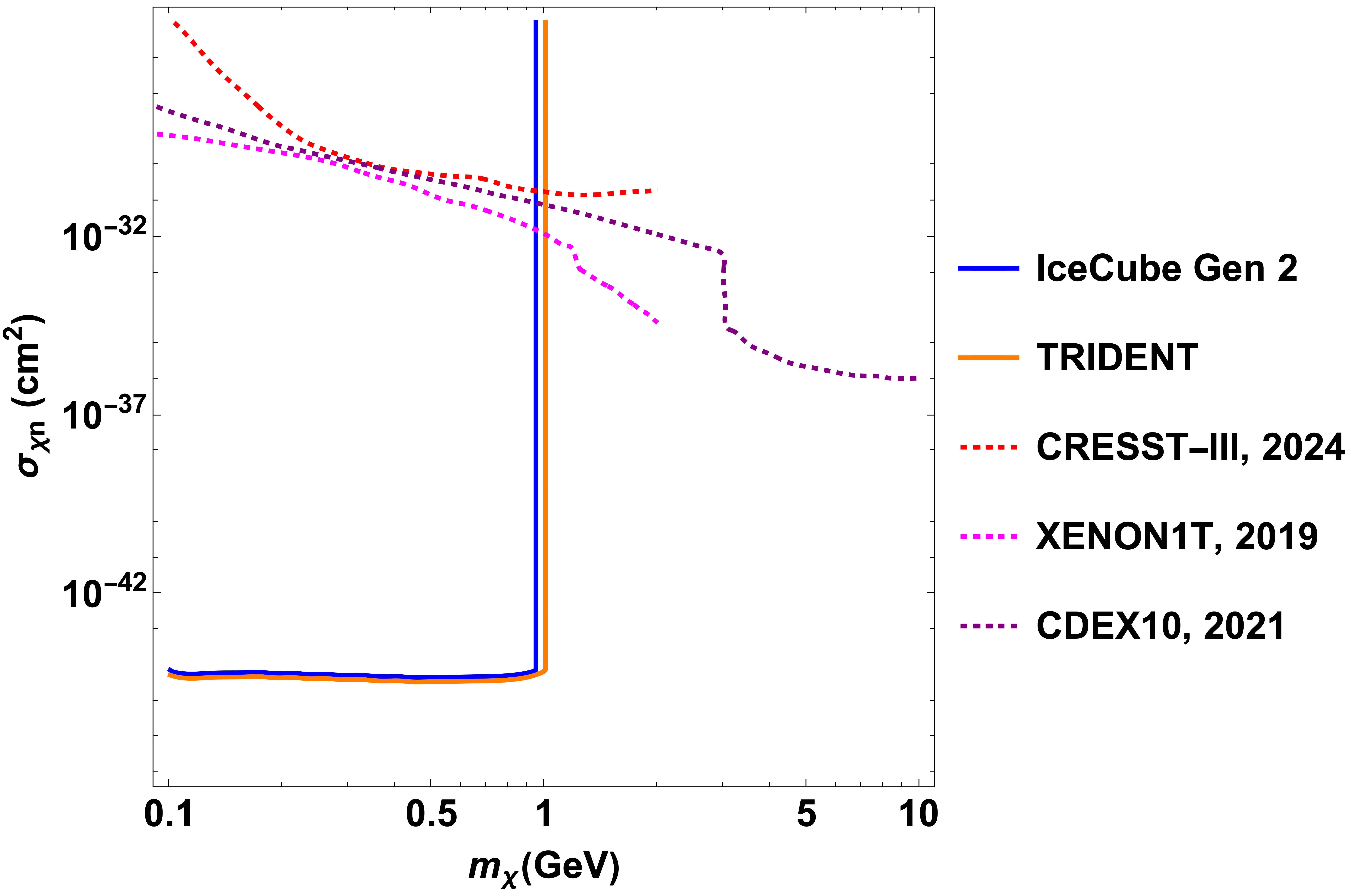}
\caption{Current bounds from local NSs (solid lines) and comparison with experiments (dashed lines). } 
\label{fig:bounds_current}
\end{figure}

\noindent We consider the DM annihilation process via long-lived mediators as discussed in Sec.\ref{sec:formulation}, and DM annihilation into only one specific channel with 100\% branching fraction of a species (neutrino or gamma-ray). As a result, we obtain the maximum upper limit on the capture rate which is then translated into the DM scattering cross section.\\

\noindent Figure~\ref{fig:bounds_current}, with conservative measures, we translate the differential flux sensitivity (Fig. \ref{fig:diff_flux_dec}) at $\delta = 0^0$, of our selected neutrino detectors at Sec.~\ref{sec:flux_dec} and derive the stacked constraints\footnote{For obtaining the stacked flux values, we combine the expected flux from each NS. This treatment generally applies to improve the sensitivity, especially for the null or faint detection.} on $\sigma_{\chi n}$ as a function of the DM mass, $m_\chi$ by combining the flux limits from 8 local NSs (Eqs.~(\ref{eqn:dm_flux_gamma}) and (\ref{eqn:dm_flux_nu})). The limits shown in Fig.~\ref{fig:bounds_current} emerge from the intricate interplay of the DM capture rate, DM density profile, and DM mass range and serve as the current constraints on $\sigma_{\chi n}$ for 8 nearby local NSs. We derive the stacked limits for the NFW density profile as mentioned in Sec.~\ref{sec:dm_capture}. To ensure the robustness of our findings, we validate the results against those obtained for the cored Burkert profile \cite{Burkert:1995yz, Salucci:2000ps}. However, our analysis reveals $\sigma_{\chi n}$ has no significant dependence on the density profile for local NSs. \\

\noindent Figure\ref{fig:bounds_current} highlights that with 8 local NSs, we only obtain the limits for DM mass $<$ 1 GeV, and only IceCube Gen 2 and TRIDENT can provide any meaningful bounds. Here we want to stress that as for Fig.\ref{fig:bounds_current}, we currently get the limits only for DM mass $<$ 1 GeV, we correct the DM capture rate by implementing the Pauli Blocking effect (Eq.~(\ref{PB_capture})). Furthermore, in Fig.~\ref{fig:bounds_current}, we compare the limits with experimental bounds obtained from direct detection bounds, such as CRESST-III\cite{CRESST:2024cpr} with a total exposure of 3.64 kg-day, CDEX10\cite{CDEX:2021cll} with 205.4 kg-day of exposure and XENON1T\cite{XENON:2019zpr} for one tonne-year of exposure. For DM mass $<$ 1 GeV, a handful of nearby NSs provide stronger bounds than obtained from XENON1T~\cite{XENON:2019zpr}. \\

\noindent The observations of Fig.\ref{fig:bounds_current} is consistent with the finding of Fig.~\ref{fig:diff_flux_dec}. An NS with a certain maximum capture rate will have an emission of maximum flux (neutrino or gamma-ray) due to the annihilation of DM. If the DM annihilation flux from the local NS remains below the sensitivity of the detector, it will remain undetected and no constraints on DM mass and its scattering cross section can be placed.\\

\subsubsection{Limits from the local population of NSs with neutrino and gamma-ray telescopes}
\label{sec:limits_pop_gamma_nu}

\begin{figure}[htbp]
\centering
\includegraphics[width=0.64\textwidth]{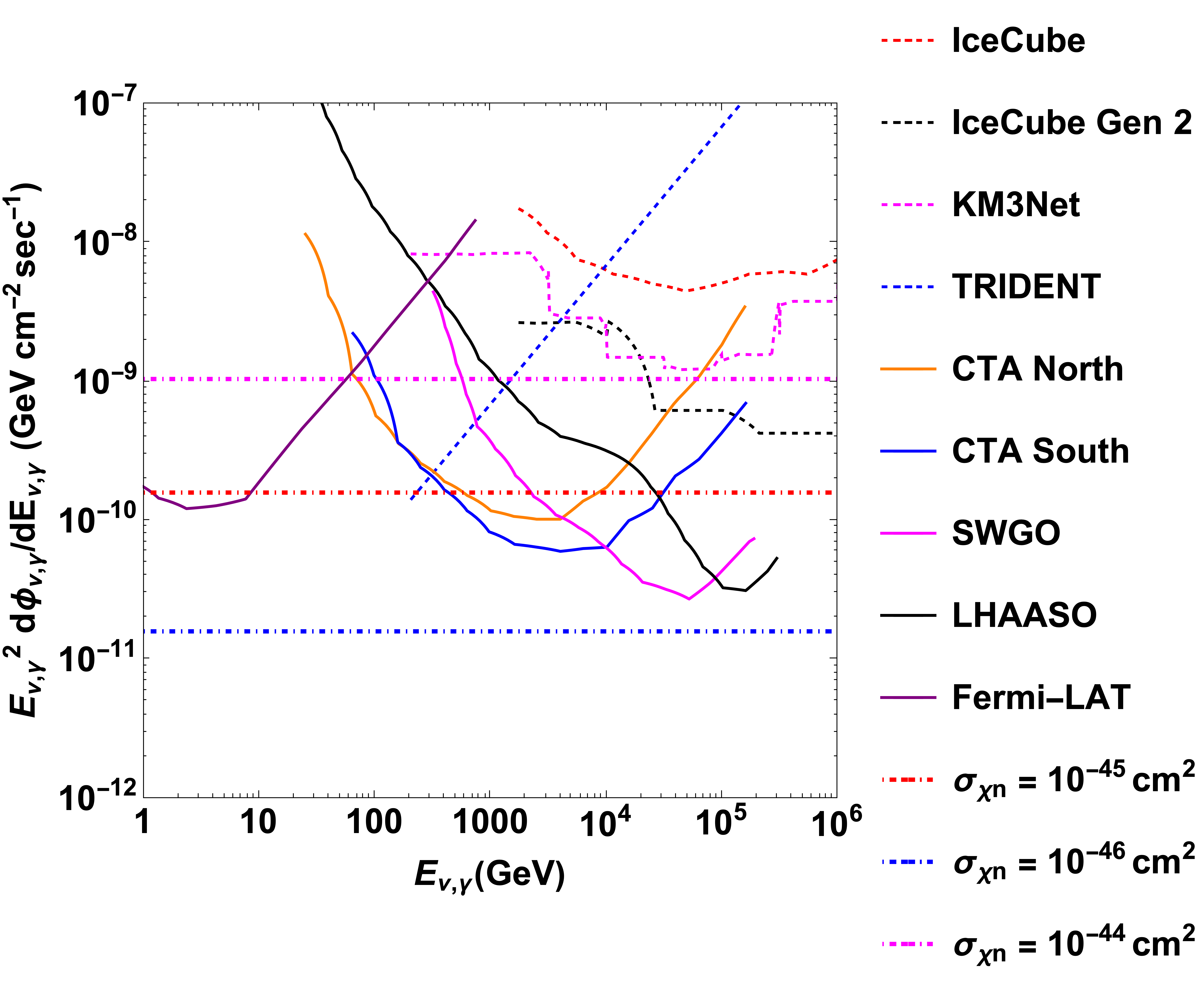}
\caption{DM annihilation spectra through long-lived mediators for three scattering cross section values expected from the local NS population with $N_{\rm NS}~\approx~10^{5}$ within 0.6 kpc from Earth.}
\label{fig:dm_spectra}
\end{figure}

\noindent In Sec.~\ref{sec:limits_nu_only}, we observe that the set of local NSs considered in Sec.\ref{ns_obs} fails to provide any conclusive limit upon DM capture for GeV scale DM. This also indicates that the bound obtained from only a few locally detected NSs is less significant from the statistical point of view and also less detectable with the present sensitivity of the detectors. However, the population study of local NSs with a distribution based on the DM halo profile and CGW searches has the potential to improve the present significantly reported limits on DM capture rate for GeV scale DM.\\

\noindent Following the population model proposed by Ref. \cite{Reed:2021scb}, we anticipate 
$N_{\rm NS}~\approx~10^{5}$ within 0.6 kpc from Earth for NS ellipticity $\epsilon~\sim~10^{-5}$. In Fig.~\ref{fig:dm_spectra}, we present the expected differential gamma-ray and neutrino flux (horizontal lines) from $10^{5}$ the number of local NSs for the NFW halo profile. We use Eqs.~(\ref{eqn:dm_flux_gamma})-(\ref{eqn:dm_flux_nu}) to obtain the expected flux for energy ranging between $1 - 10^{6}$ GeV for three scattering cross section ($\sigma_{\chi n}$) values. For a chosen value of $\sigma_{\chi n}$, both gamma-ray and neutrino spectra result in the same flux limits due to box-shaped annihilation spectra. We further compare the upper limit of differential flux with the current and projected sensitivities of the gamma-ray and neutrino detectors (from Fig.~\ref{fig:diff_flux_ref}) as shown in Fig.~\ref{fig:dm_spectra}. We observe that CTA, LHASSO and SGWO is capable of probing DM annihilation flux from the local NS population for $\sigma_{\chi n}\leq 10^{-45}$ cm$^2$ in the mass range (or energy range, for boxlike spectra, they are equivalent) $m_{\chi}\geq 100$ GeV while Fermi-LAT and TRIDENT can probe for $m_{\chi}\leq 100$ GeV. The projected sensitivity of IceCube Gen 2 is promising for probing $m_{\chi}\geq 10^{5}$ GeV for $\sigma_{\chi n}~=~ 10^{-44}-10^{-45}$ cm$^2$. Moreover, we also find that IceCube and KM3Net are not sufficiently sensitive for $\sigma_{\chi n}~\leq~ 10^{-44}$ cm$^2$. In conclusion, Fig.~\ref{fig:dm_spectra} further indicates that although most experimental limits are sensitive to detection of a high annihilation flux with $\sigma_{\chi n}\geq 10^{-45}$ cm$^2$, a lower DM scattering cross section $\sigma_{\chi n}<10^{-46}$ cm$^2$ is unreachable with the current detector sensitivity with respect to the upper limit for $10^{5}$ number of local NS population from CGW searches.\\

\begin{figure}[htbp]
\centering
\includegraphics[width=0.48\textwidth]{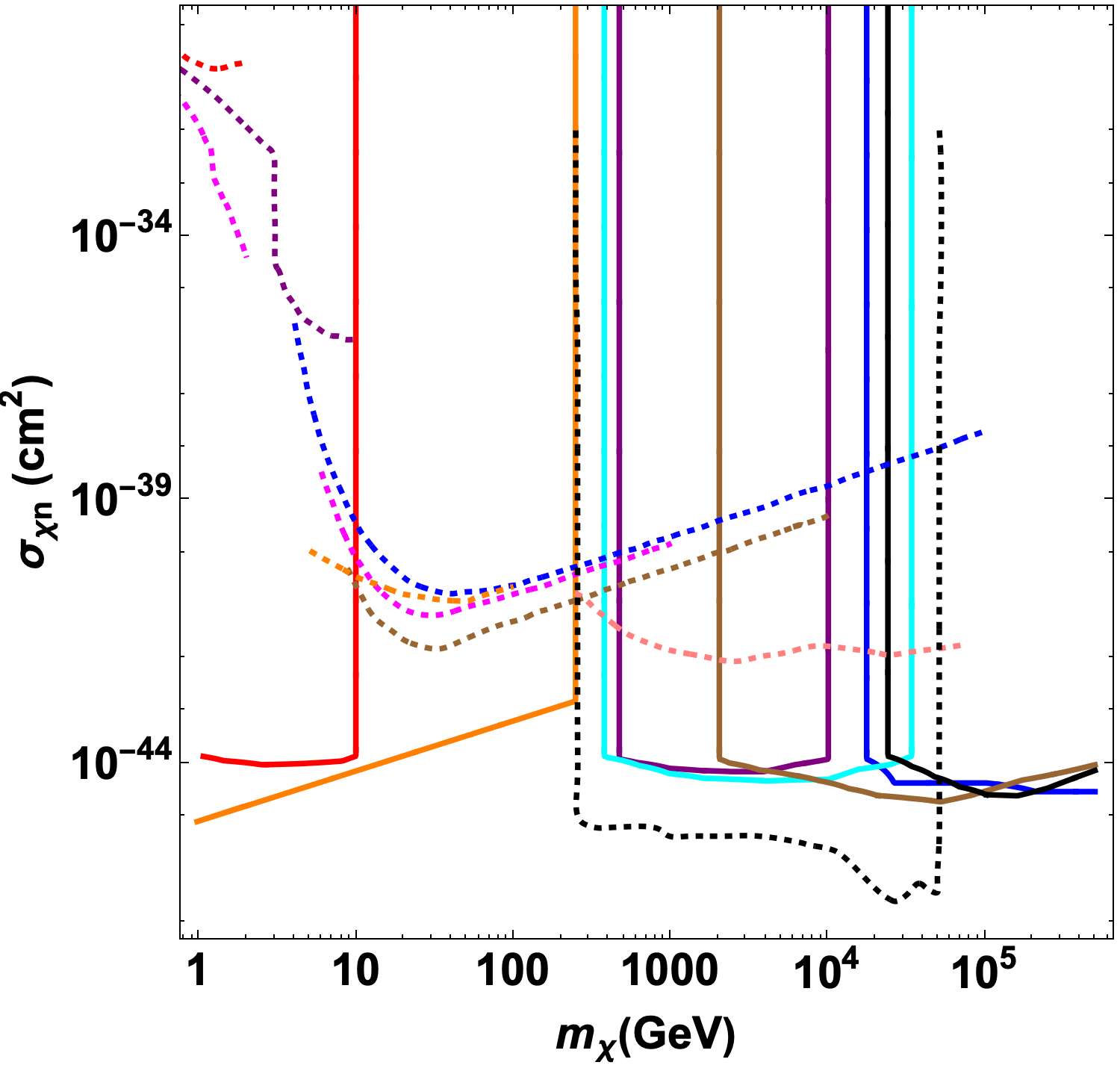}
\includegraphics[width=0.19\textwidth]{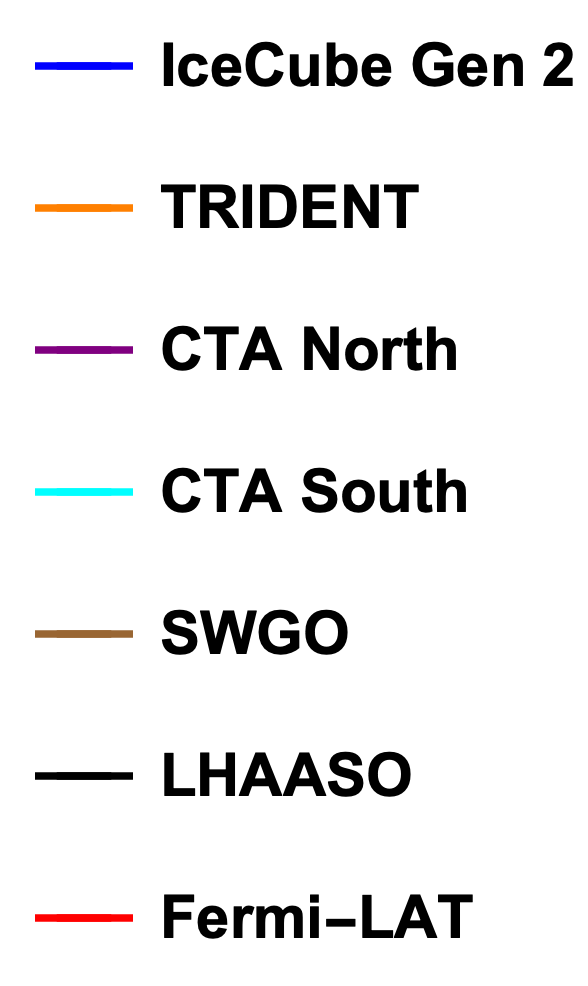}
\includegraphics[width=0.305\textwidth]{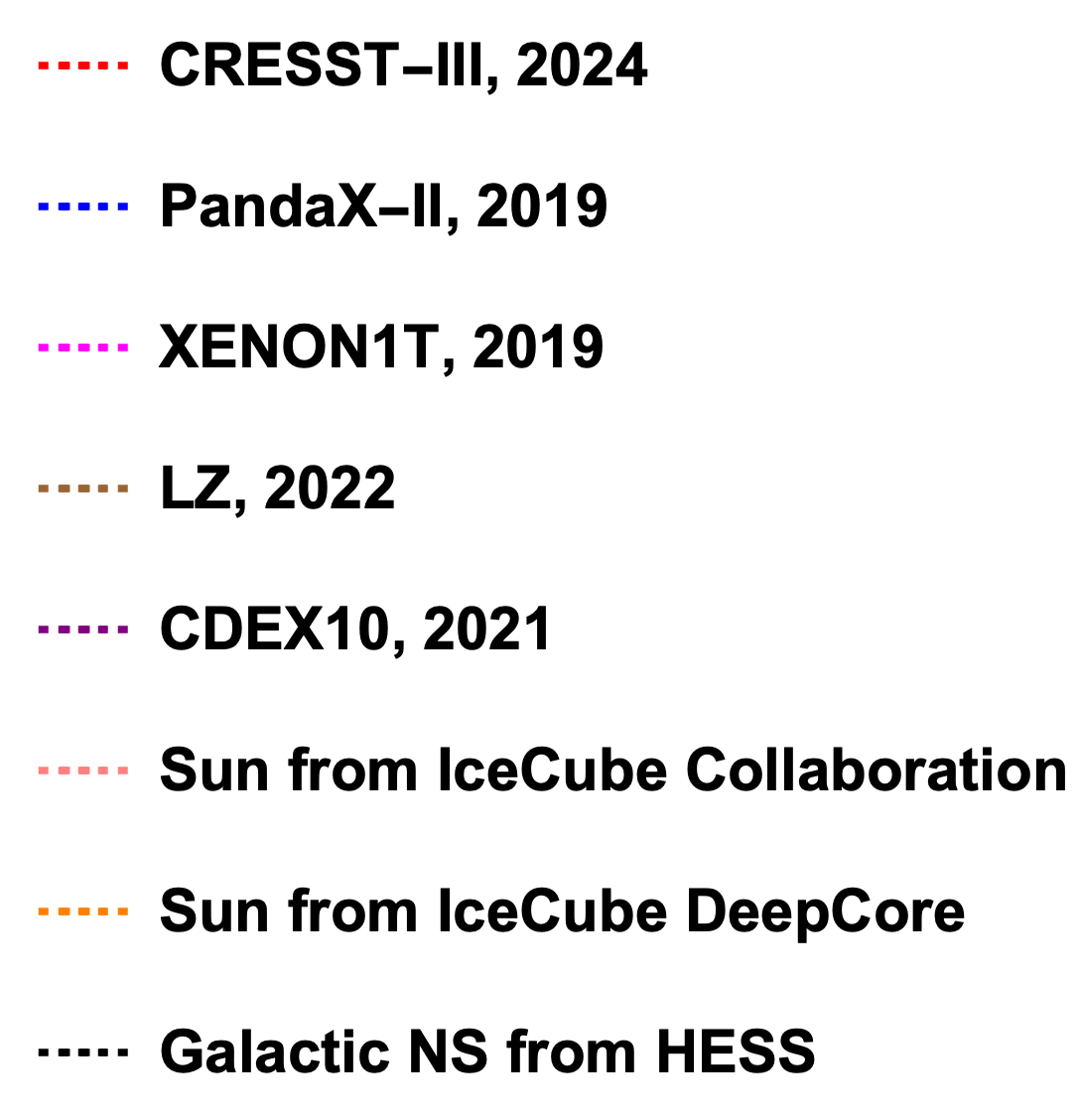}
\caption{Comprehensive limits on $\sigma_{\chi n}$ from the local NS population with CGW searches (solid lines) and comparison with several DM direct detection experiments and limits obtained from astrophysical studies.}
\label{fig:bounds_GW_exp_comp}
\end{figure}

\noindent In Fig.\,\ref{fig:bounds_GW_exp_comp}, we convert the differential flux sensitivity from Fig.\,\ref{fig:diff_flux_ref} to show the projected bounds for $10^{5}$ local NSs. In Fig.\,\ref{fig:bounds_current}, due to the limited statistics from just eight NSs, we adopt a conservative approach by using the differential flux sensitivity from Fig.\,\ref{fig:diff_flux_dec} at $\delta = 0^\circ$. For Fig.\,\ref{fig:bounds_GW_exp_comp}, where we expect to improve our bounds with the larger local NS population, we use the detector sensitivity from Fig.\,\ref{fig:diff_flux_ref}, which varies as a function of energy. Comparing with Fig.\,\ref{fig:bounds_current}, we find that the bounds predicted by the CGW search with the local NS population provide more stringent limits on the DM scattering cross section across a wide range of DM mass, from a few GeV to $10^{6}$ GeV. We restrict ourselves to the DM mass, $10^{6}$ GeV to minimize the multiple scattering aspects of DM with nucleon for the heavy DM mass \cite{Bell:2020jou}.\\

\noindent In Fig. \ref{fig:bounds_GW_exp_comp}, we further extend our analysis to compare our findings with the results obtained from various direct detection experiments. Specifically, we consider CDEX10\cite{CDEX:2021cll} with 205.4 kg-day, CRESST-III\cite{CRESST:2024cpr} with a total exposure of 3.64 kg-day, LZ\cite{LZ:2022lsv} with 60 live days, PandaX-II\cite{PandaX-II:2018woa} with a total exposure of 54-ton-days,  and XENON1T\cite{XENON:2019zpr, XENON:2019rxp} using one tonne-year of exposure corresponds to DM-neutron SD scattering cross section. We also compare our results with astrophysical limits from IceCube for Sun \cite{IceCube:2021xzo,refId0} and indirect search bounds from H.E.S.S. for Galactic population of NS \cite{Leane:2021ihh}. This comparison underscores the advantage of constraints derived from local NSs, which extend to DM masses $\sim~10^{6}$ GeV.\\

\noindent From Fig.~\ref{fig:bounds_GW_exp_comp}, we observe that at low mass regime (between 1-100 GeV), only Fermi-LAT and TRIDENT provide the bounds on $\sigma_{\chi n}$ based on their projected sensitivity, with TRIDENT imposing the strongest limit, i.e., $\sigma_{\chi n}~\sim~10^{-45}$ cm$^2$, which is several orders of magnitude stronger than the present direct detection and astrophysical bounds. Similarly, in the heavy DM regime ($m_{\chi}\geq 1000$ GeV), we obtain similar bounds on $\sigma_{\chi n}$ from searches of the local population of NSs with the projected sensitivity of CTA, IceCube Gen 2, and LHAASO, while SWGO imposes the most stringent limit of $\sigma_{\chi n}~\sim~10^{-45}$ cm$^2$, surpassing direct detection experiments by a considerable margin. Among all astrophysical bounds compared in Fig.~\ref{fig:bounds_GW_exp_comp}, H.E.S.S. bound from Galactic NS contribution is on par with our limits on $\sigma_{\chi n}$. Therefore, Fig.~\ref{fig:bounds_GW_exp_comp} suggests that CGW missions with improved search capabilities for local NSs have the potential to tighten the limits on the DM scattering cross section across a wide range of DM masses, which we are generally unable to probe with a few detected local NSs.\\

\section{Conclusion}
\label{sec:conclusion}

\noindent Probing NSs in light of DM capture offers valuable insights into the nature of this elusive substance, shedding light on its properties and interactions with ordinary matter. This research holds promise for advancing our understanding of fundamental physics and unlocking the mysteries of the cosmos. Our motivation for this work is to target the nearby local NSs and study the DM capture rate inside them. Generally, NSs are considered to be very efficient in capturing DM with feeble interactions because of their small radius and strong lower gravitational focus power. But, to this date, only a handful number of NSs have been detected (even though they are expected to be very abundant in the galaxy $\approx~10^{8}$) and most of them are not very close to Earth (as compared to BDs). This eventually also weakens the bounds on scattering cross section ($\sigma_{\chi n}$) when our goal is to detect the gamma rays and neutrinos resulting from DM annihilation with space or ground-based detectors. Recently, gravitational wave searches have gained a lot of attention, fueling our interest in the detection of local NSs, being the natural source of CGW emission. The search for CGW indicates a substantially rich NS population adjacent to Earth and strengthens the bound on $\sigma_{\chi n}$. For the purpose of this paper, we combine the sensitivity from current and future generation gamma-ray and neutrino detectors which enables us to construct a comprehensive picture of DM signature from local NSs.
\\

\noindent The conclusion of this article is as follows:
\begin{itemize}
    \item Our investigation delves into the extensive analysis of the sensitivity at the 90\% C.L. and the potential for 5$\sigma$ discovery across present and future-generation neutrino telescopes. We specifically examine their capability to detect neutrino signals originating from our selected NSs resulting from the decay of long-lived mediators produced through DM annihilation (see Fig.~\ref{fig:diff_flux_dec}). Among the current neutrino detectors, only TRIDENT exhibits the potential to detect neutrinos for DM masses around 1 GeV with a scattering cross section ($\sigma_{\chi n}$) of $6 \times 10^{-45}$ cm$^{2}$. However, for DM masses $m_{\chi} \geq$ 5 GeV, none of the existing neutrino telescopes are adequate unless there are brighter sources or a further enhancement in DM-nucleon scattering.
    
    \item We interpret the stacked constraints on $\sigma_{\chi n}$ concerning $m_\chi$ by combining the flux limits from 8 nearby NSs (Fig. \ref{fig:bounds_current}) assuming an NFW density profile. Fig. \ref{fig:bounds_current} indicates that constraints are only obtainable for DM masses $<$ 1 GeV with 8 local NSs, and meaningful bounds can be provided by IceCube Gen 2, and TRIDENT.

    \item We further explore the potential impact of the expected population of local NSs from CGW searches on our current limits (Fig.~\ref{fig:dm_spectra}). With the unknown NS population taken into account from CGW searches by Reed \ie \cite{Reed:2021scb}, we conclude that present gamma-ray and neutrino detectors are sensitive to GeV scale DM with $\sigma_{\chi n}\simeq 10^{-45}$ cm$^2$.

    \item We derive the bounds on $\sigma_{\chi n}$ from the cumulative emission of the local NS population within 0.6 kpc from Earth (Fig.~\ref{fig:bounds_GW_exp_comp}). The cumulative emission expected from approximately $10^{5}$ NSs indicates that the most stringent bounds of $\sigma_{\chi n}~\sim~10^{-45}$ cm$^2$ originates from TRIDENT (for DM annihilation into neutrinos) for DM masses $m_{\chi}<$ 1000 GeV and from SWGO (for DM annihilation into gamma-rays) for DM masses $m_{\chi}>$ 1000 GeV. The bounds obtained on $\sigma_{\chi_n}$ are found to be stronger than the IceCube limits from the Sun and a few orders of magnitude weaker than the H.E.S.S. limit obtained from the Galactic NS population.

\end{itemize}    
\section{Discussion}
\label{sec:Discussion}
\noindent This study upholds a promising avenue for local NS from a data analysis perspective with some limitations that can impact our results. We also want to add that when searching for a DM signal from a local neutron star, the impact of backgrounds for neutrinos and gamma rays is less significant compared to diffuse regions. For gamma rays, the proximity of the source improves the signal-to-noise ratio, while energy and angular selection suppress astrophysical backgrounds. For neutrinos, directional constraints and reduced atmospheric backgrounds at higher energies aid in isolating the signal. Following we discuss some possible limitations on our current results.\\

\noindent In this study, we adopted a simplified model using a zero-temperature approximation irrespective of the NS's core temperature.
This is a general set-up applied by many literatures to provide bound on DM properties from NS~\cite{Leane:2024bvh, Acevedo:2024ttq, Leane:2021ihh, Bose:2021yhz}. While higher core temperatures are likely, especially in younger stars, this approach allows us to focus on the primary objectives without introducing the complexities of temperature-dependent factors. Although incorporating the actual core temperature could yield more precise results, it would require additional assumptions and models beyond the scope of our work. We acknowledge that our results may vary by a few orders with improved temperature corrections imposed on NSs leaving room for more detailed future work. Furthermore, in this study, we have opted to use a simplified model where the key parameters, such as the mass of NS, escape velocity, and nucleon form factors, are treated as constants within the NS interior. This approach is intended to provide a manageable framework that offers clearer insights into the core dynamics of our model, allowing for direct comparisons against previous studies for other celestial objects. However, we acknowledge that these quantities vary significantly across the NS interior due to gravitational effects and changes in density, which can influence the capture rate as noted in refs. \cite{Bell:2020jou, Bell:2020obw, Anzuini:2021lnv}. We also want to point out that, for this work, we use the standardized values for $M_{\rm NS}$ and $R_{\rm NS}$, but the capture rates directly depend on these two parameters. For neutron star mass $M_{\rm NS}$ between 1.4 $M_{\odot}$-2.5 $M_{\odot}$ and $R_{\rm NS}$ from 10-25 km, our reported bounds could vary by nearly one order of magnitude. We are aware of the limitations of this method and plan to address these complexities in future work to more accurately reflect the true nature of NS environments. \\

\noindent With the present formalism, we find a significant bound on the DM scattering cross section from the local NS population. Improved constraints on the $\sigma_{\chi _n}$ can be achieved with precise corrections to the NS properties and the nature of DM interactions. The limits on $\sigma_{\chi n}$ obtained for DM mass $m_{\chi}$ are valid for DM with specific scalar-scalar type interactions. Including different DM interaction types will change the relevant bounds in the present work and a detailed study is required. One can also consider the case where DM undergoes direct annihilation instead of annihilation via a long-lived mediator considered in the present work. However, this involves different dynamics of final state annihilation with avenues that can be addressed in future works.\\

\noindent In addition, we want to point out that the precision of our findings strongly hinges on the systematic uncertainties associated with each instrument. Regarding gamma-ray detectors, the Fermi-LAT reveals flux uncertainties of approximately 10\%-15\%, alongside variations in energy scale (up to 5\%-10\%) and effective area (around 10\%) \cite{Ackermann_2012, Abdo_2009}. CTA typically experiences flux uncertainties spanning 10\%-20\%, accompanied by calibration uncertainties of 5\%-10\% and atmospheric influences ranging from 10\%-15\% \cite{CTAConsortium:2017dvg, Bernlohr:2008kv}. LHAASO estimates flux uncertainties at 15\%-20\%, with calibration around 10\% and atmospheric modeling impacting results by 10\%-15\% \cite{DiSciascio:2016rgi, Wang:2023qeg}. SWGO's flux uncertainties are approximately 15\%-20\%, with calibration and atmospheric effects each contributing about 10\%-15\% \cite{Albert:2019afb}. Concerning neutrino detectors, IceCube displays flux uncertainties ranging from 10\%-20\%, with optical properties of ice introducing uncertainties of up to 10\%-15\% and calibration uncertainties around 10\% \cite{Williams:2020mvu, IceCube:2016zyt}. IceCube Gen 2 demonstrates similar uncertainties, slightly improved to around 10\%-15\% \cite{IceCube-Gen2:2020qha, IceCube:2014gqr}. KM3Net reports flux uncertainties of 15\%-20\%, with uncertainties in water properties and calibration reaching up to 10\%-15\% each \cite{KM3Net:2016zxf, LeBreton:2019lpq}. TRIDENT anticipates flux uncertainties of around 15\%-20\%, with calibration and environmental conditions contributing approximately 10\%-15\% each \cite{Ye:2023dch, Li:2023wqk}. In summary, systematic uncertainties in our chosen detectors may influence our results by approximately 10\%-20\%.\\

\noindent In conclusion, this study reveals that present and upcoming neutrino and gamma-ray observatories are not very sensitive to DM-nucleon scattering when a few known local NSs within 0.6 kpc in the neighborhood of Earth are considered and can only provide a weak bound for DM mass $\sim 1$ GeV. Improved limits on DM-nucleon scattering cross section can be achieved with the consideration of about $10^{5}$ local NS population anticipated from the CGW search for a range between low ($\sim$ sub-GeV) to heavy ($\sim$ PeV) DM mass. The corresponding limit on $\sigma_{\chi n}$ with the local NS population is commendable concerning those obtained from DM direct search and astrophysical limits from the Sun and convincing as we compare with indirect bounds from the Galactic NS population. Future CGW searches with better sensitivity are expected to limit further the local NS population, which can further constrain the DM-nucleon scattering with better precision.\\

\acknowledgments
\noindent A. D. B. acknowledges financial support from DST, India, under Grant No. IFA20-PH250 (INSPIRE Faculty Award). Some part of P. B.’s work was supported by the EOSC Future project which was cofunded by the European Union Horizon Program call INFRAEOSC-03-2020, Grant Agreement No. 101017536. P. B. further acknowledges support from the COFUND action of Horizon Europe’s Marie Sklodowska-Curie Actions research program, Grant Agreement 101081355 (SMASH). The authors would like to thank Dr. Tarak Maity, Dr. Nicholas L. Rodd, and Dr. Sandra Robles for their valuable suggestions.

\bibliography{main_prd_published}
\end{document}